\def\breakon{\end{multicols}\widetext\vspace{.5cm}
\noindent\rule{.48\linewidth}{.3mm}\rule{.3mm}{.5cm}\vspace{.5cm}}
\def\breakoff{\vspace{.5cm}
\noindent
\rule{.52\linewidth}{.0mm}\rule[-.47cm]{.3mm}{.5cm}\rule{.48\linewidth}{.3mm}
\vspace{.5cm}
\begin{multicols}{2}
\narrowtext}
\def\ie{{\it i.\ e.\/}}
\documentstyle[aps,epsf,multicol,eqsecnum]{revtex}
\begin{document}
\title{Effective field theory for the bulk and edge states of quantum Hall states in
unpolarized single layer and bilayer systems}
\author{Ana Lopez$^1$ and Eduardo Fradkin$^2$}
\address{$^1$Centro At{\'o}mico Bariloche, 8400 S.\ C.\ de Bariloche,
R{\'\i}o Negro, Argentina\\ $^2$Department of Physics, University
of Illinois at Urbana-\-Champaign, 1110 West Green Street, Urbana,
IL 61801-3080}
\bigskip
\date{\today}
\maketitle

\begin{abstract}
We present an effective theory for the bulk Fractional Quantum
Hall states in spin-polarized bilayer and spin-$1/2$ single layer
two-dimensional electron gases (2DEG) in high magnetic fields
consistent with the requirement of global gauge invariance on
systems with periodic boundary conditions.
We derive the theory for the edge states that
follows naturally from this bulk theory. We find that the minimal
effective theory contains two propagating edge modes that carry charge
and energy, and two non-propagating topological modes responsible
for the statistics of the excitations. We give a detailed
description of the effective theory for the spin-singlet states,
the symmetric bilayer states and for the $(m,m,m)$ states. We
calculate explicitly, for a number of cases of interest, the
operators that create the elementary excitations, their bound
states, and the electron. We also discuss
the scaling behavior of the tunneling conductances in different
situations: internal tunneling, tunneling between identical edges
and tunneling into a FQH state from a Fermi liquid.
\end{abstract}
\bigskip

\begin{multicols}{2}
\narrowtext

\section{Introduction}
\label{sec:int}

Two-dimensional electron gases (2DEG) in bilayers in high magnetic fields
display a rich spectrum of fractional quantum Hall (FQH) states,
some of them  exhibiting rather unusual properties not found in
single layer systems. Likewise unpolarized and partially polarized
single layer FQH states have properties related to the FQH states
in fully polarized bilayer states.  Wave functions for the
$(m,m,n)$ primary, Laughlin-like, FQH states in bilayers were
proposed long ago by Halperin\cite{halperin-mmn}. A description of
these states in terms of the hierarchy was given by Wen and
Zee\cite{wenzee-matrix}, and a fermion Chern-Simons
construction\cite{paper1} of their Jain-like states were given by
us in ref.\ [\ref{ref:paper2}] and [\ref{ref:capitulo}]. The basic
properties of the bulk $(m,m,n)$ Halperin states are well
understood, both experimentally and theoretically, for a review
see ref.\ [\ref{ref:review}]. The bulk Halperin states are well
described by the effective $K$-matrix
theories\cite{wenzee-matrix}\cite{frohlich}, and their associated
edge states have a simple representation in terms of a theory of
two chiral bosons\cite{wen}.

Among the multitude of bilayer states, there are three
representative cases of special interest:
\begin{enumerate}
\item
The spin-singlet states in single layers. These states have an
$SU(2)$ spin symmetry that organizes their
spectra\cite{paper2,FB}. The most interesting representatives are
the $(3,3,2)$ state which corresponds to a $\nu=2/5$ singlet
state, and the related state at $\nu=2/3$, which belongs to the reversed sequence of the $(3,3,1)$ state
with $p<0$. Both states have been seen experimentally
\cite{eisenstein-23,singlet-23}. More recently higher order
descendents in the Jain-like family of spin singlet states at
$\nu=4/7$ and $\nu=4/9$ have also been seen\cite{cho47}.
\item
The symmetric bilayer states. These states have a $U(1) \times
U(1)$ symmetry. The most interesting representative of this group
is the $(3,3,1)$ FQH state in bilayers at $\nu=1/2$. This state
has a number of unusual properties and it has been see
experimentally\cite{eisenstein-331}. It has been conjectured that
the $(3,3,1)$ state may have a phase transition to the non-Abelian
Pfaffian FQH state\cite{moore-read,gww} triggered by inter-layer
tunneling\cite{halperin-diagram}. Experiments on wide quantum
wells\cite{wide} show clear evidence of other bilayer states, including the
$(3,3,1)$ descendants at $4/5$ and $4/7$. They also show strong evidence
unbalanced states (at zero parallel field) at $11/15$, $13/21$ among others.
\item
The $(m,m,m)$ bilayer states. These states are special in that
they are ground states of the 2DEG with spontaneous  interlayer
coherence\cite{wenzee-gapless}\cite{ezawa}\cite{indiana}.
Consequently, their spectra have a massless bosonic excitation, a
Goldstone boson. Recent tunneling experiments into the bulk
$(1,1,1)$ state at filling factor $1$ have provided evidence for
the Goldstone boson\cite{spielman}. There is also a related state
at filling factor $1$ in single layers which exhibits both the
integer quantum hall effect and
ferromagnetism\cite{sondhi-kivelson}\cite{indiana}. NMR
experiments\cite{barrett} have provided firm evidence for the
existence of skyrmion excitations in these
states\cite{sondhi-kivelson}.
\end{enumerate}

Recent experiments on tunneling of electrons into the edge of (so
far) single layer fully polarized FQH states by A.\ Chang and
coworkers\cite{chang1,chang2} have opened a new window to study
these interesting systems. No experiments have been done yet on
bilayer systems or in unpolarized single layer 2DEGs. The
experiments have raised a number of questions that theory has not
 answered so far. It has been found that in clean junctions the
tunneling current at low voltages does follow a scaling law, as
predicted by theory\cite{wen-tunnel}\cite{kane-fisher}. The
exponent found in experiment obeys, to a good approximation, the
law $\alpha=1/\nu$ where $\nu$ is the filling factor. There is an
indication in the experiment of a small feature that may be
interpreted as a plateau in the exponent near
$\nu=1/3$\cite{chang2}. However, this interpretation requires a
number of assumptions about the charge distribution near the edge
which are hard to justify in detail. While other interpretations
are also possible, the clearest conclusion that can be drawn from
the data is that the dependence of the exponent on the magnetic
field is much simpler than predicted by theory. At the theoretical
level there are two types of predictions. On the one hand, Kane
and Fisher\cite{kane-fisher}, using the standard hierarchy
construction, predicted a highly complex number theoretic function
of the filling factor in the absence of disorder, and a
significantly simpler dependence when the effects of disorder are
accounted for\cite{kane-fisher,KFP}. Shytov, Levitov and
Halperin\cite{shytov} have considered the effects of long range
Coulomb interactions and charge redistribution on edge tunneling
and found a result identical to the case of short range
interactions with disorder. However both predictions disagree
quite significantly with the experimental results. Not only the
predicted structure of the dependence of the exponent with the
filling factor is much more complex than what is observed, but
even the average tendency of this dependence fails to predict a
simple $\alpha=1/\nu$ law. In fact, for $\frac{1}{3}\leq \nu \leq
\frac{1}{2}$ the prediction is a {\sl constant} value of
$\alpha=3$.

This puzzle has motivated the development of a number of alternative approaches to this problem.
D.\ H.\ Lee and X.\ G.\ Wen \cite{wen-lee} have shown that if the
velocity of the neutral modes of the standard $K$-matrix description
for some reason becomes much smaller than the velocity of the charge mode, the
$\alpha=1/\nu$ law follows. Z{\"u}licke and Macdonald have
suggested that the experiment could be understood if it were a
simple consequence of charge conservation\cite{zulicke}. (Note
however that the electron operator used in \cite{zulicke} has the
correct charge but the wrong statistics; however this shortcoming
 does not affect the tunneling density of states.)

Recently we have proposed a conceptually different construction of
the edge states which also yields the $\alpha=1/\nu$
law\cite{ref1}. Our approach is based on the observation that
 the complex structure of the edge states predicted by the
$K$-matrix theory\cite{wen} is a reflection on the edge of the
standard hierarchical construction of the bulk FQH state, and that
a possible, and consistent, interpretation of the experiments is
that they fail to show evidence for the hierarchy. The principal
idea behind the work of ref.\ [\ref{ref:ref1}] is that, for all
the states in the Jain sequence, there is a much simpler
construction of their bulk effective theories which does not
involve the full structure of the hierarchy-based $K$-matrix theory,
and the (sometimes) large number of additional conservation laws
that it requires. The resulting effective theory contains one
charge field and at least one additional field needed to give the
correct topological degeneracies on a torus, and the correct
quantum numbers for the excitations. One advantage of this
approach is that, unlike the standard hierarchical construction,
this theory  has only one fundamental quasiparticle from which the
entire spectrum is built.

In ref.\ [\ref{ref:ref1}] we also showed that there is a natural
edge associated to this picture, which holds under the assumption
that the edge is sharp and that no edge reconstruction takes
place. The assumption that the edge is sharp means that the fine
structure of edge modes, predicted by the mean field approximation
to the fermion Chern Simons theory of the bulk
state\cite{chklovskii}, occurs only on small scales (\ie\ the
magnetic length). At long scales these modes {\sl cannot be
resolved} (even assuming that the mean field approximation can be
trusted in regions where the gap collapses, something that is far
from clear). In this regime, the effective theory of the edge
acquires a minimal structure which contains one propagating
charged mode and, (at least) one extra mode required by the
topological invariance of the bulk. This latter mode, which in
this paper we will call the {\sl topological} mode, has no
relation with the neutral modes that appear in Wen's theory
\cite{wen}. In fact this mode {\sl does not} propagate simply
because there is no physical mechanism to give the excitations of
this field a non-zero velocity. Thus, this field remains
topological at the edge: its Hamiltonian is zero. In practice, the
only role of the topological field is to contribute to the
operators that create the excitations in the form of effective
Klein operators that set the correct statistics. In this way, the
operators of this effective theory have the right quantum numbers.
A direct consequence is that the predicted tunneling density of
states for electrons has the correct exponent and naturally yields
the $\alpha=1/\nu$ law. However, this construction can only
describe the system exactly at the values of the filling fraction
for the Jain sequence. In other words, it can not make any
prediction about the tunneling exponents for filling fractions
interpolating between two FQHE states.

The $\alpha=1/\nu$ law of single layer fully polarized systems
suggests that this effect may be a simple consequence of charge
conservation (as suggested in ref.\ [\ref{ref:zulicke}]). If this
assumption were to be correct, a similar prediction should also
hold for more complex situations such as in the FQH states in
bilayers, spin unresolved single layer systems, and other cases
where more that one subband may be present\cite{jiang}, or for 2D
hole gases where Kramers states are present\cite{coleridge}. A few
predictions for tunneling exponents for the simplest Halperin
states have been proposed in the literature\cite{tunnel-bilayers}.
However, a theory that accounts for the rich spectrum of
excitations and the related tunneling processes in these FQH
states is lacking. In particular, a general theory of the
tunneling exponents has not yet been formulated. (see however
ref.\ \ref{ref:chamon-moore}).

In this paper we present a theory of the edge states for all
Jain-like states of FQH states in bilayers and unpolarized single
layer systems.  We  generalize the approach of our earlier work
for single layer fully polarized systems\cite{ref1}. We begin by
constructing the simplest possible effective theory of the {\sl
bulk} states, compatible with the requirements of global gauge
invariance, topological degeneracy on a torus, and with the
smallest possible number of fundamental quasiparticles. We then
use this theory to determine the physics of its edge states. Here
too, under the assumption that the edge is sharp, unreconstructed
and clean, we proceed to derive the simplest theory compatible
with these requirements. The effective theory of the edge states
thus derived contains two propagating fields, the charge modes for
each layer, and two topological non-propagating fields. We then
use this construction to determine the structure of the edge
states for all Jain sequences whose primary states are the
Halperin $(m_1,m_2,n)$ states. We discuss in detail the spectrum
of operators for the symmetric states. In particular we discuss
the theory of the edge states for the $SU(2)$ states, based on the
$(m,m,m-1)$ states and the general $U(1) \times U(1)$ states. In
each case we give an explicit construction of the operators that
create the quasiparticles (and quasiholes), charged and neutral
bound states (including neutral fermionic states) and the electron
operators. For the case of the $SU(2)$ states we show how the
symmetry is realized in the spectrum and  how it is promoted to a
local $su(2)_1$ current algebra. We use these theories of the
edges to calculate the tunneling exponents for all cases in three
different situations: internal tunneling, tunneling of electrons
between identical liquids and tunneling of electrons into a FQH
fluid from an external Fermi liquid lead. We find that although
the tunneling exponents are universal, in general they no longer follow
the $1/\nu$ law observed in single layer fully polarized systems.

Finally we  present a theory of the edge states for the $(m,m,m)$
states. Here we discuss the role of the bulk Goldstone boson and
its effects on edge physics. The effective theory of the edges of
the $(m,m,m)$ states is a chiral boson for the charge mode (with
the same radius as the Laughlin states), and a non local
non-chiral theory for the neutral modes. The non-locality of the
neutral sector is due to the massless Goldstone mode. The
effective action for the neutral sector is a generalization of the
Caldeira-Leggett action to a problem of a field interacting with
an active medium.

\subsection{Summary of Results and Organization}

This paper is organized as follows. In Section \ref{sec:bl} we
derive an effective theory for bilayer FQH states in the bulk,
consistent with the requirement of global gauge invariance on a
torus. We apply this construction for general Jain-like states on
bilayers. We find that a generic state is labeled by five integers, 
$n_1$, $n_2$, $p_1$, $p_2$ and $n$. Here, $2n_1$ and $2n_2$ are the 
number of flux quanta attached to a fermion on
layers $1$ and $2$ respectively, $n$ is the number of flux quanta attached 
to a fermion on layer $1$  due to a fermion on layer $2$ (and viceversa), 
and $p_1$ and $p_2$ are the occupation
numbers of the effective Landau levels. The signs of $p_1$ and $p_2$ determine
deterimne both direct and reversed sequences of FQH states. The formulas for the total filling
fraction $\nu=\nu_1+\nu_2$ and polarization ${\cal M}=(\nu_1-\nu_2)/2\nu$ are given by
Eqs.\ \ref{eq:nu} and \ref{eq:polarization}. These sequencies were derived by us in
reference \cite{paper2}. In general the states in these sequencies describe polarized
states, namely states with ${\cal M} \neq 0$. The symmetric states, with $2n_1=2n_2=m-1$,
with $m$ an odd integer, play a special role. In particular for $m=n+1$ these states have
an effective $SU(2)$ spin symmetry. Otherwise the symmetry is $U(1) \times U(1)$.  

We derive an effective
$K$-matrix theory for a general bilayer state, and discuss its implications for the
symmetric states, both $SU(2)$ (where the layer index is regarded
as the spin of the electron) and $U(1) \times U(1)$. 
In both cases we discuss the structure of the excitation spectrum. 
Unlike the conventional $K$-matrix,
based on the Haldane-Halperin hierarchical construction, here the rank of the 
$K$-matrix is the same for all states in the sequencies, and the entries of the matrix are
parametrized by the five integers $n_1$, $n_2$, $p_1$, $p_2$ and $n$. The spectrum
contains only two fundamental quasiparticles (or quasihole), and the remaining states are
constructed as bound states.
These results generalize our recent work on single layer fully polarized systems, 
\cite{ref1}.
We also derive an effective theory for the
$(m,m,m)$ states, and discuss the nature of the excitations, including a construction of
the electron and of the quasiparticles. 

In Section \ref{sec:edge} we derive the
theory of the edge states for bilayers, under the assumptions that
the edge is sharp, clean and unreconstructed. In Section
\ref{sec:op}  we construct the operators that create the
quasiparticles, their bound states, and the electron at the edge. We derive explicit
expressions of these operators for both the $SU(2)$ and $U(1) \times U(1)$ symmetric
states. Here we compute their propagators and find the tunneling exponents
for different physical settings of particular interest. 
In particular we use these results to derive scaling laws for internal tunneling 
of quasiparticles, and for tunneling of electrons both between identical FQH states (on
these sequencies) and from a Fermi liquid into a state on these sequencies. We find that,
unlike the case of a single layer with fully polarized electrons,
in this case the exponent $\alpha$ of the tunneling current versus voltage {\sl is not}
simply determined by the filling fraction $\nu$. 

\begin{table}
\begin{center}
  \begin{tabular} {|c|c|c|c|}\hline
  $\nu$ &  $\alpha_{qp}$    &     $\alpha_e$  & $\alpha_t$    \\   \hline \hline
  ${\displaystyle{\frac{2p}{2np+1}}}$    &     ${\displaystyle{\frac{p(2n-2np-1)+2}
  {2np+1}}}$ for $p>0$               &  $2n-1+{\displaystyle{\frac{2}{p}}}$       & 
  $n+{\displaystyle{\frac{1}{p}}}$         \\
            &         ${\displaystyle{\frac{(2n-2n|p|+1)}{2n|p|-1}}}$ for $p<0$   &  $2n-1$  &  $n$ \\    \hline
            $2/5$  &  $1/5 $   &   $5$  &   $3$     \\  \hline
            $2/3$  &  $1/3 $   &   $3$  &   $2$     \\  \hline
            $4/9$  &  $-4/9$   &   $4$  &   $5/2$   \\  \hline
            $4/7$  &  $1/7 $   &   $3$  &   $2$   \\  \hline
  \end{tabular}
  \caption{Tunneling exponents for $SU(2)$ bilayer states.}
\label{table:table1}
\end{center}
\end{table} 
In Tables \ref{table:table1} and \ref{table:table2} we summarize
these results and give a few examples for states that have been observed experimentally.
Recall that for the $SU(2)$ symmetric bilayer states the integer $n$ is even, and $p$ 
is an arbitrary integer. We have
listed the exponents for internal tunneling of quasiparticles on the left
column on Table \ref{table:table1}. On the center column we give the exponents for electron tunneling between
identical states and on the right column we give the exponents for tunneling of electrons from a Fermi liquid into
an $SU(2)$ FQH state. In the text we also give tunneling exponents for bound states in several filling
fractions.In Table \ref{table:table1}, the states at $2/5$ and $4/9$ and the reversed sequence states at $2/3$
and $4/7$. As a general rule these filling factors normally appear in mire than one sequence. For instance $4/7$ can
be realized either as a singlet state (as we have just seen) or as a member of a $U(1) \times U(1)$ sequence, as we
will see next. These are distinct states and both can be realized in a given bilayer system. By varying a parameter
(such as pressure or density) it should be possible to induce phase transitions among these states. 

\breakon
\begin{table}
\begin{center}
  \begin{tabular} {|c|c|c|c|}\hline
  $\nu$ &  $\alpha_{qp}$    &     $\alpha_e$  & $\alpha_t$    \\   \hline\hline
  ${{2p}\over{(m+n-1)p+1}}$    &     ${2({m-1+1/p})\over {((m-1)p+1)^2-(np)^2}}-1$ for $s>0$ &  $2(m-1+1/p)-1$       &  $m-1+1/p$         \\
            &         ${{2n}\over {(np)^2-((m-1)p+1)^2}}-1$ for $s<0$   &    &   \\    \hline
            $2\over{m+n} $  &  ${2m\over{m^2-n^2}}-1$ &     $2m-1$ &   $m$   \\  \hline
            $ 1/2$   & $-1/4 $    &    $5$  &   $3$     \\ \hline
            $ 4/7$   & $ -16/21$  &    $4$  &   $5/2$   \\ \hline
            $ 4/5$   & $  -2/5 $  &    $2$  &   $3/2$   \\ \hline
           \end{tabular}
	   \caption{Tunneling exponents for $U(1) \otimes U(1)$ bilayer states.}
  \label{table:table2}
\end{center}
\end{table}
\breakoff
\noindent
In the case of the $U(1) \times U(1)$, $m$ and $n$ are odd integers, $s={\rm sign}( m-n -1 + 1/p)$, and $p$ is 
an arbitrary integer. Here, $4/7$ is the first descendant in the series of the primary Halperin state $(3,3,1)$.
It has $m=3$, $n=1$, $p=2$ and $s>0$. The state at $4/5$ is the reversed state with $p=-2$. LIkewise,
the unbalanced states observed in wide quantum wells at $\nu=11/15$ (with $\Delta \nu/\nu=\pm 1/11$) and at $\nu=13/21$
(with $\Delta \nu/\nu=\pm 1/13$) are also descendants of the $(3,3,1)$ state with $(p_1,p_2)=(-5,-3)$ and $(-3,-5)$ and
$(p_1,p_2)=(3,7)$ and $(7,3)$ respectively.  

Finally,
in Section \ref{sec:mmm-edge} we discuss in detail the special
case of the $(m,m,m)$ states. Here we show  how the properties
of their edge states are affected by the existence of a gapless
(Goldstone) mode in the bulk. In particular we give an explict derivation of the theory
of the edge states. We find that the galpless bulk states yield an effective theory at the
edge which is non-local. We discuss  how these features affect the electron
operator and the tunneling exponents. In Section \ref{sec:conclusions} we discuss our
conclusions and open questions. In Appendix \ref{app:AA} we compute
some operator product expansions needed to find the physical
operators at the edges.

\section{Bilayer systems}
\label{sec:bl}

In this section we construct an effective bulk theory for the 2DEG
in strong magnetic fields in bilayer systems. We can follow here
exactly the same point of view that we presented  in references
\cite{ref1}\cite{chetan} for single-layer fully polarized 2DEG,
with the difference that now there is an extra degree of freedom
represented by the layer index. Although the basic philosophy is
the same, the physics of these systems is different.

We begin with a first quantized Feynman path-integral of the 2DEG.
In the case of bilayers the worldlines of the particles can be
represented by a conserved 3-current $j_\mu^\alpha$. The index
$\alpha=1,2$ labels the layer where the particle moves. From now
on, whenever the layer indices are between brackets it will be
understood that the indices are not summed over. The weight of a
particular history of the 2DEG in the Feynman path-integral is
given by the action of the history. This action contains terms
representing the kinetic and interaction energies of the 2DEG. In
addition, since the particles are fermions there is a minus sign
whenever any pair of particles are exchanged. We will assume that
the individual histories are such that the paths do not cross. In
other words, there are short range interactions that will keep the
particles apart. Hence, relative to a ``parent configuration" in
which there are no exchanges, configurations with exchanges can be
viewed as a set of {\sl linked} paths. The sign associated with a
given configuration is then $(-1)^{N_P}$, where the number of
permutations $N_P$ is equal to the linking number of the
configuration.

We now proceed with the flux-attachement transformation, which
maps fermions to fermions. The linking number of the trajectories
of the particles is given by
\begin{equation}
\nu_L^{\alpha \beta}[j_\mu^{(\alpha)}]= \int d^3x \; j_\mu^\alpha
(x) b_\mu^\beta(x) \label{jdotBD}
\end{equation}
with $j_\mu^\alpha (x)= \epsilon_{\mu \nu \lambda} \partial^\nu
b_\lambda^\alpha(x)$, where $\alpha,\beta=1,2$ are layer indices.
If the world lines of the particles do not cross, then
 $\nu_L^{\alpha \beta}[j_\mu]$ is a topological invariant. Clearly, for a
bilayer system, the linking number $\nu_L^{\alpha \beta}$ is a
symmetric matrix. Naturaly, the same observation holds for
particles with spin.

Thus, if $S[j_\mu^\alpha]$ is the action for a given history, the
quantum mechanical amplitudes of all physical observables remain
unchanged if the action is modified by
\begin{equation}
S[j_\mu^\alpha] \to S[j_\mu^\alpha]-2\pi  T^{\alpha \beta}
\nu_L^{\alpha \beta}[j_\mu^{\alpha}] \label{shiftB}
\end{equation}
with
\begin{equation}
2T^{\alpha \beta}= \left(
\begin{array}{cc}
2n_1        & {n}\\ n & 2n_2
\end{array}
\right) \label{eq:T}
\end{equation}
where $n_1,n_2$ and $n$ are arbitrary integers. The condition on
the off-diagonal elements comes from the fact that $T^{\alpha
\beta}$ is necessarily a symmetric matrix since the linking number
is  symmetric as well.

Following the same approach introduced in references
\cite{ref1,chetan} we find that, for a bilayer system, the action
is given by
\begin{equation}
S_{\rm eff}[a,b,j]= {\frac{1}{2\pi}} a_\mu^\alpha \epsilon_{\mu
\nu \lambda} \partial^\nu b_\lambda^\alpha- a_\mu^\alpha
j_\mu^\alpha- {\frac{2T^{\alpha \beta}}{4\pi}} \epsilon_{\mu \nu
\lambda} b_\mu^\alpha
\partial^\nu b_\lambda^\beta \label{eq:effectiveB}
\end{equation}
The indices $\alpha, \beta=1,2$ label the layer or spin
polarization, while $\mu,\nu,\lambda$ are the space-time indices.
If the currents $j_\mu^a$ correspond to the electrons living in
the layer $\alpha$ of the system, and since we assume that there
is no tunneling between layers, the effective action written in
second quantization becomes, in the composite fermion picture
\begin{eqnarray}
S_{\rm eff}=&& \int d^3x \; \left(\psi^\dagger_\alpha (i
D_0^{\alpha}+\mu^\alpha)\psi_\alpha -{\frac{1}{2M}} |D_j^{\alpha}
\psi_\alpha|^2 \right) \nonumber\\
&&\!\!\!\!\!\!\!\!\!\!\!\!\!\!\!\!\!\!\!\! -{\frac{1}{2}} \!
\!\int \!\!\!d^3x \!\!\int \!\!\! d^3x' (|\psi_\alpha(x)|^2-{\bar
\rho}_\alpha) V_{\alpha \beta}(|{\vec x}-{\vec x}'|)
(|\psi_\beta(x')|^2-{\bar \rho}_\beta) \nonumber\\ +&&\int d^3x
\left[{\frac{1}{2\pi}} a_\mu^\alpha \epsilon_{\mu \nu \lambda}
\partial^\nu b_\lambda^\alpha - {\frac{2T^{\alpha \beta}}{4\pi}}
\epsilon_{\mu \nu \lambda} b_\mu^\alpha \partial^\nu
b_\lambda^\beta \right] \nonumber\\ && \label{eq:Seffbilayer}
\end{eqnarray}
where $D_\mu^\alpha=\partial_\mu-i(e A_\mu^\alpha+a_\mu^\alpha)$,
$A_\mu^\alpha$ is the external electromagnetic field acting on
layer $\alpha$, and $\mu^\alpha$ is the chemical potential for
layer $\alpha$.

The mean field theory of this problem, {\it i.\ e.\/} the average
field approximation, is found by requiring that the effective
action $S_{\rm eff}$ of Eq.\ \ref{eq:Seffbilayer} be stationary,
\begin{equation}
{\frac{\delta S_{\rm eff}}{\delta a_\mu^\alpha(x)}}=0 \; ,\qquad
{\frac{\delta S_{\rm eff}}{\delta b_\mu^\alpha(x)}}=0
\label{eq:SPA}
\end{equation}
for $\alpha=1,2$. For fluid states these equations become,
\begin{equation}
{\bar \rho}_\alpha=-{\frac{1}{2\pi}} \langle\epsilon_{ij}
\partial_i \ b_j^\alpha \rangle\; , \qquad \; \langle\epsilon_{ij}
\partial_i a_j^\alpha \rangle = 2 T^{\alpha \beta}
\langle\epsilon_{ij} \partial_i  b_j^\beta \rangle \label{eq:SPA2}
\end{equation}
where ${\bar \rho}_\alpha$ is the charge density of the fermions.

It is straightforward to show that these equations are identical
to the mean field equations of our earlier work on bilayer FQH
states~\cite{paper2}. Hence, we find the same series of fractions
for the bilayer system and, for special choices of interactions,
for single layer systems with unpolarized or partially polarized
spins. For the bilayer system the allowed fractions are
\begin{eqnarray}
\nu_1 &=& {\displaystyle \frac{n-\left(\pm
{\frac{1}{p_2}}+2n_2\right)} {n^2-\left(\pm
{\frac{1}{p_1}}+2n_1\right) \left(\pm
{\frac{1}{p_2}}+2n_2\right)}} \nonumber \\ \nu_2 &=&
{\displaystyle \frac{n-\left(\pm {\frac{1}{p_1}}+2n_1\right)}
{n^2-\left(\pm {\frac{1}{p_1}}+2n_1\right) \left(\pm
{\frac{1}{p_2}}+2n_2\right)}} \nonumber \\ && \label{eq:fractions}
\end{eqnarray}
Following the methods of reference \cite{paper2}, we expand the
effective action around the mean field solution with uniform
density. We find an effective Lagrangian ${\cal L}_{\rm eff}$ at
long distances and low energies, of the form (including the
coupling to a weak external gauge field $A_\mu^a$)
\begin{eqnarray}
{\cal L}_{\rm eff}[a,b,e] =&&
{\frac{p_a}{4\pi}}\epsilon_{\mu\nu\lambda} a_{\mu}^\alpha
\partial_{\nu} a_{\lambda}^\alpha  +{\frac{1}{2 \pi}}
\epsilon_{\mu\nu\lambda} a_{\mu}^\alpha \partial_{\nu}
b_{\lambda}^\alpha \nonumber\\
 && - {\frac{2T^{\alpha,\beta}}{4 \pi}}\epsilon_{\mu\nu\lambda}
b_{\mu}^\alpha \partial_{\nu} b_{\lambda}^\beta -{\frac{1}{2 \pi}}
\epsilon_{\mu\nu\lambda} A_{\mu}^\alpha
\partial_{\nu} b_{\lambda}^\alpha
\nonumber\\ && +{\frac{1}{4\pi}} \epsilon_{\mu\nu\lambda} e_{\mu}
\partial_{\nu} e_{\lambda} - a_{\mu}^\alpha
j_{qp,a}^{\mu} -  e_{\mu} ( j_{qp,1}^{\mu}+j_{qp,2}^{\mu})
\nonumber\\ &&
 \label{eq:a3B}
\end{eqnarray}
where $p_\alpha$ ($\alpha=1,2$) are arbitrary integers (positive
or negative). As usual, the sequences of allowed FQH states can have either sign of $p_\alpha$. We will refer to
the sequences with  $p_\alpha>0$ as the direct sequence and those with $p_\alpha<0$ as the reversed sequence. 
Intuitively, the sign of $p_\alpha$ is the sign of the effective magnetic field felt by the composite fermions,
relative to the  sign of the external field that acts on the electrons.

The current $j_{qp,\alpha}^{\mu}$ represents the quasiparticles.
Since in the fermionic picture the bare quasiparticles are
composite fermions, the statistics of all the excitations that we
will compute is defined relative to fermions. Therefore we have
introduced  the field $e_{\mu}$ to keep track of the underlying
statistics, {\it i.\ e.\/} to transform the fermions into bosons.
Notice that the quasiparticle current couples directly to the
gauge fields $a_\mu^\alpha$ and $e_\mu$ but not to the
hydrodynamic field $b_\mu^\alpha$ or to the external
electromagnetic field.

We can write the effective Lagrangian in a more compact form by
defining, in the  basis $(b_\mu^1, a_\mu^1, b_\mu^2, a_\mu^2,
e_\mu)$, the ($2 \times 5$) charge matrix
\begin{equation}
t_{aI}= \left(
\begin{array}{ccccc}
1 & 0 & 0 & 0 & 0 \\ 0 & 0 & 1 & 0 & 0
\end{array}
\right) \label{eq:t}
\end{equation}
as well as a $2 \times 5$ quasiparticle coupling matrix
$\ell_{\alpha I}$. The set of allowed quasiparticle states is
represented by matrices $\ell_{\alpha I}$ of the form
\begin{equation}
\ell_{\alpha I}= \left(
\begin{array}{ccccc}
 0 & k_1 & 0   & 0 & -k_1 \\
 0 & 0   & 0 & k_2& -k_2
\end{array}
\right) \label{eq:ells}
\end{equation}
 where $k_1$ and $k_2$ are arbitrary integers.
In addition, we also define the ($ 5 \times 5$) Chern-Simons
coupling constant matrix $K$ of the form
\begin{equation}
K= \left(
\begin{array}{ccccc}
-2n_1 &  1    & -n    & 0   &  0  \\
 1    &  p_1  &  0    & 0   &  0  \\
-n    &  0    & -2n_2 & 1   &  0  \\
 0    &  0    &  1    & p_2 &  0  \\
 0    &  0    &  0    & 0   &  1
\end{array}
\right) \label{eq:Keff}
\end{equation}
whose determinant is given by
\begin{equation}
\det K= (1+ 2n_1 p_1) (1 + 2n_2 p_2) - n^2 p_1 p_2
\end{equation}
It is well known that the determinant of the $K$ matrix is equal
to the degeneracy of the ground state of the FQH fluid on a torus
\cite{hosotani1,wen-topo,wen-niu,hosotani2}.

It is easy to show  that the only possible solutions for
 $\det K =0$ are given by $p_1=p_2=1$ and $2n_1 +1 =2n_2 +1 =
n=m$ which correspond to the $(m,m,m)$ states. We  describe now
the  states such that $\det K \neq 0$ and come back to the
$(m,m,m)$ states later.

With the above definitions, the Lagrangian of the effective theory
of the FQHE in bilayers takes the form
\begin{equation}
{\cal L}= {\frac{1}{4\pi}}K_{IJ}\epsilon_{\mu \nu \lambda}
a_I^\mu\partial^\nu a_J^\lambda - {\frac{1}{2\pi}} t_{\alpha
I}A_\alpha^\mu \epsilon_{\mu \nu \lambda} \partial^\nu
\alpha^\lambda_I+
 \ell_{\alpha I} j^{\mu}_{qp,\alpha}
a_I^\mu \label{eq:eff9}
\end{equation}
where $I,J=1,\ldots,5$ and $\alpha=1,2$. There are two types of
quasiparticles, given by the two choices of the coupling vectors
$\ell_I$ of Eq.\ \ref{eq:ells}. The coupling to external
electromagnetic perturbations is given by the charge matrix $t^T$
of Eq.\ \ref{eq:t}. Notice however that, even though the structure
of the effective Lagrangian has the form required by the general
classification of Abelian FQH states of Wen and Zee
\cite{wenzee-matrix}, the physical interpretation of the component
fields is actually quite different from those of the standard
hierarchy.

The $ 2 \times 2$ Hall conductance matrix for these states is
found to be equal to
\begin{eqnarray}
\sigma_{xy}^{\alpha \beta} =&&{\frac {1}{2\pi}} t_{\alpha I}
K^{-1}_{IJ} t_{J \beta} \nonumber\\ =&& {\frac{1}{2\pi \det K }}
\left(
\begin{array}{cc}
 p_1 ( 2n_2 p_2 +1) & -np_1 p_2         \\
-np_1 p_2           & p_2(2n_1 p_1 +1)
\end{array}
\right) \label{eq:ff}
\end{eqnarray}
The total conductance of the bilayer system is $\sigma_{xy}^{\rm
total}={\frac{\nu}{2\pi}}$ for total filling factors of the form
\begin{equation}
\nu=\sum_{\alpha,\beta} \nu_{\alpha,\beta}= {\frac{p_1 ( 2n_2 p_2
+1)+p_2(2n_1 p_1 +1)-2np_1 p_2}{\det K }} 
\label{eq:nu}
\end{equation}
which agrees with our earlier work\cite{paper2}. The filling
factor of each layer is $\nu_\alpha=\sum_\beta
\nu_{\alpha,\beta}$. Let us define the polarization per particle
${\cal M}_{\rm total}$ of the ground state of the system as
\begin{equation}
{\cal M}_{\rm total} \equiv{\frac{1}{2}}
 \left({\frac{N_1-N_2}{N_1+N_2}}\right)
= {\frac{\nu_1-\nu_2}{2\nu}}
\label{eq:polarization}
\end{equation}
For non-polarized and partially polarized systems, in which the
layer index $\alpha$ is the spin index of the electrons, this
definition of the polarization coincides with the total
$z$-component {\it magnetization} per electron of the ground state
$S^z_{\rm total}$. From Eq.\ \ref{eq:nu} we find that ${\cal
M}_{\rm total}$ is given by
\begin{equation}
{\cal M}_{\rm total}={\frac{1}{2}} \left(
{\frac{(p_1-p_2)-2(n_1-n_2)p_1p_2} {p_1 ( 2n_2 p_2 +1)+p_2(2n_1
p_1 +1)-2np_1 p_2}} \right) \label{eq:M}
\end{equation}
The fundamental vortices (or quasiholes) are represented by the
coupling matrices of Eq.\ \ref{eq:ells} with the choices
$(k_1,k_2)=(1,0), (0,1)$. Let us label by $a=1,2$ these
quasiparticles. For the  quasihole $\alpha$, the charge matrix
$Q^{bc}_{qp,a}$, in units of the electric charge $e$,
 is  given by ($b,c=1,2$)
\begin{eqnarray}
Q_{bc}^{qp,1}=&& t_{bI} K^{-1}_{IJ} \ell_{Jc}^1= {\frac{1}{\det
K}} \left(
\begin{array}{cc}
2n_2p_2 +1 &   0   \\
 - np_2    &   0
\end{array}
\right) \nonumber \\ Q_{bc}^{qp,2}=&& t_{bI} K^{-1}_{IJ}
\ell_{Jc}^2= {\frac{1}{\det K}} \left(
\begin{array}{cc}
0   &   - np_1   \\ 0   & 2n_1p_1 +1
\end{array}
\right) \nonumber \\ && \label{eq:Qs}
\end{eqnarray}
Each quasiparticle can be characterized by a charge $Q_{qp,a}$ and
a polarization $M^a_{qp}$. The charges and polarizations have the
form
\begin{eqnarray}
Q_{qp,a}=&&\sum_{bc} Q^{bc}_{qp,a} \\
M_{qp,a}=&&-{\frac{1}{2}}\sum_{bc} (-1)^c Q^{bc}_{qp,a} \equiv
{\frac{1}{2}} Q^s_{qp,a} \label{eq:QM}
\end{eqnarray}
Notice that we have defined the quasiparticle polarization as
measured in units of the quasiparticle charge, and with the factor
of ${\frac{1}{2}}$ it is equal to the $z$-component of the spin of
the quasiparticle (or quasihole). Alternatively, we may use the
``spin charge" $Q^s_{qp,a}$ as defined in the last line of Eq.\
\ref{eq:QM}, which agrees with the notation of Milovanovic and
Read\cite{milovanovic}.

Therefore the quasihole charges and polarizations are given by
\begin{eqnarray}
 Q_{qp}^{1}=&&
  {\frac {2n_2p_2+1 - np_2 }{\det K }} = {\frac {\nu_1}{p_1}}
\nonumber \\ M_{qp}^{1}=&&{\frac{1}{2}} {\frac {2n_2p_2+1 + np_2
}{\det K }} \nonumber \\ Q_{qp}^{2}=&& {\frac {2n_1p_1 +1 - np_1
}{\det K }}={\frac{\nu_2}{p_2}} \nonumber \\
M_{qp}^{2}=&&-{\frac{1}{2}} {\frac {2n_1p_1 +1 + np_1 }{\det K }}
\nonumber \\ && \label{eq:qpch}
\end{eqnarray}
The statistics of the quasiholes, measured relative to bosons, is
given by the matrix $\theta^{ab}_{qp}$ which, in general, has the
form
\begin{equation}
{\frac{\theta^{ab}_{qp}}{\pi}}= \ell_{aI} K^{-1}_{IJ} \ell_{Jb}
\end{equation}
The statistics of the elementary quasiparticles are found to be
\begin{eqnarray}
{\frac{\theta^{1}_{qp}}{\pi}}=&&{\frac{\theta^{11}_{qp}}{\pi}} =1+
{\frac{2n_1 + p_2 (4n_1n_2 - n^2)}{\det K}} \nonumber \\
{\frac{\theta^{2}_{qp}}{\pi}}=&&{\frac{\theta^{22}_{qp}}{\pi}} =1+
{\frac{2n_2 + p_1 (4n_1n_2 - n^2)}{\det K}} \nonumber \\
{\frac{\theta^{12}_{qp}}{\pi}}=&& 1+ {\frac{n}{\det K}} \nonumber
\\ && \label{eq:qpst}
\end{eqnarray}
Let us consider now the charge and statistics of a composite
object made of $k_1$ quasiparticles of type $1$ and $k_2$
quasiparticles of type $2$, with $k_1$, $k_2$ integers. The total
charges and polarizations of these composite objects, in units of
the electron charge $e$, are
\begin{eqnarray}
Q &&= {\frac {k_1(2n_2p_2+1 - np_2)+k_2(2n_1p_1 +1 - np_1 ) }{\det
K }} \nonumber \\ M &&= {\frac {k_1(2n_2p_2+1 + np_2)-k_2(2n_1p_1
+1 + np_1) } {2 \det K}} \nonumber \\ && \label{eq:totQM}
\end{eqnarray}
and their statistics is given by
\begin{eqnarray}
\lefteqn{{\frac{\theta}{\pi}}=
 (k_1+k_2)^2} &&\nonumber \\
&& + {\frac { 2n_1 k_1^2 + 2n_2 k_2^2+ 2 n k_1 k_2 + (p_2 k_1^2+
p_1 k_2^2) (4n_1n_2 - n^2)}{\det K}}\nonumber \\ &&
\label{eq:statcomp}
\end{eqnarray}
It is easy to check that if we take $k_1=2n_1p_1+1$ quasiparticles
of type $1$ and $k_2=np_2$ quasiparticles of type $2$, the
resulting operator has $Q=-e$, $M=1/2$ and fermionic statistics
since, in this case
\begin{eqnarray}
{\frac {\theta}{\pi}}=  (2n_1p_1+1)&&(2n_1(p_1+1)+1) \nonumber \\
&&+n^2 p_2 (p_2+1)+ 2np_2(2n_1p_1+1) \nonumber \\ &&
\label{eq:theta-odd}
\end{eqnarray}
is always an odd integer.

In the next section  will discuss the implications of these
general results in the context of several cases of  interest. In
section \ref {sec:op} we will see how to combine quasiparticle
operators to get the correct electron operator.

\subsection{$SU(2)$ symmetry and Spin Singlet States}
\label{subsec:singlet}

We will now regard the two layers as a spin index with $\alpha=1$
being spin $\uparrow$ and $\alpha=2$ being spin $\downarrow$. For
these special cases, in order to preserve the $SU(2)$ spin rotation invariance,
we must choose $2n_i=m-1=n$, {\it i.\ e.\/},
we do the flux attachment in such a way that it does not
distinguish between in-layer and inter-layer labels. Then we get,
as expected,  $\theta_{qp}^{\uparrow \downarrow}=
\theta_{qp}^{\uparrow}=\theta_{qp}^{\downarrow}$. Thus this
formalism can be used to describe the problem of electrons with
spin. Notice that $n$ is an even integer.

The allowed FQH states compatible with the choice $2n_i=m-1=n$,
which requires $n$ to be even, have filling fractions
\begin{equation}
\nu={\frac{p_1+p_2}{1+n(p_1+p_2)}} \label{eq:nusinglet}
\end{equation}

Of these states, only those in which $p_1+p_2$ is an even integer
are fully compatible with the $SU(2)$ symmetry. In other words, only for
$p_1+p_2$ {\sl even} it is possible to have a spin singlet state. In contrast, for
$p_1+p_2$ {\sl odd} the ground states are always polarized. The total ground
state polarization per particle $M_{\rm total}$ is the spin
polarization of the ground state $S^z_{\rm total}$, which for
these states is
\begin{equation}
S^z_{\rm total}={\frac{1-\nu n}{2\nu}} (p_1-p_2)=\frac{1}{2}
\left({\frac{p_1-p_2}{p_1+p_2}} \right) \label{eq:Szsinglets}
\end{equation}
We will classify the excitations according to their charge
$Q_{qp}$, their spin polarization $S^z_{qp}= M_{qp}$ and their
statistics $\theta_{qp}$. In all the $SU(2)$ states both
fundamental quasiholes, specified by the choices $(k_1,k_2)=
(1,0)$ and $(k_1,k_2)=(0,1)$ respectively, have the same charge
(in units where the electric charge $e=1$)
\begin{equation}
Q_{qp}^\uparrow=Q_{qp}^\downarrow=1-n \nu
\label{eq:qpcharge-singlet}
\end{equation}
and their spin polarizations $S^z_{qp}$ are, using  Eq.\
\ref{eq:qpch} with $S^z_{qp,\uparrow,\downarrow}=M_{qp,1,2}$
\begin{equation}
S^z_{qp,\uparrow,\downarrow}=\pm{\frac{1}{2}} - n\nu S^z_{\rm
total}
\end{equation}
The quasihole states have the statistics
\begin{equation}
{\frac{\theta_\uparrow}{\pi}}={\frac{\theta_\downarrow}{\pi}}=
{\frac{\theta_{\uparrow\downarrow}}{\pi}}=1+n-\nu n^2
\label{eq:qpstat-singlet}
\end{equation}
The spin singlet (or unpolarized) FQH states are given by the
choice $p_1=p_2=p$, and the filling  factors of these states are
$\nu=2p/\left(1+2np\right)$. For the spin singlet states $S^z_{\rm
total}=0$, and the elementary excitations are arranged into
$SU(2)$ multiplets. All states in a given multiplet have the same
charge and statistics but different spin polarization, as can be
read of Eqns.\ \ref{eq:qpcharge-singlet}-\ref{eq:qpstat-singlet}.
In particular, the elementary quasiholes form the doubly
degenerate (fundamental) representation of $SU(2)$ and carry the
same charge and statistics, and opposite spin polarization.

A general composite excitation is made out of $k_1$ quasiparticles
of type $1$ and $k_2$ quasiparticles of type $2$, and it has the
following quantum numbers
\begin{eqnarray}
Q_{qp}=&&(k_1+k_2) (1-n\nu) \nonumber \\
S_{qp}^z=&&{\frac{1}{2}}(k_1-k_2) \nonumber \\
{\frac{\theta}{\pi}}=&&(k_1+k_2)^2(1+n-\nu n^2) \nonumber \\ &&
\label{eq:qnssinglets}
\end{eqnarray}
Thus, the quantum numbers of the states are specified by two
integer-valued labels, $k_1$ and $k_2$, which span a two
dimensional integer lattice.

It is interesting to construct the quasi-electron states for all
the spin singlet FQH states. Since these excitations must have
$Q=1$, we must require that $k_1+k_2=1+2np$. The spin polarization
of these states is $S^z_{qp}[k_1,k_2]={\frac{1}{2}} (k_1-k_2)$.
The statistics of these states is
${\frac{\theta}{\pi}}=(1+2np)(1+2np+n)$, which is an odd integer.
Hence, these states are fermions with charge $e$. However, since
as we can see, these states actually belong to a $2np+2$-fold
degenerate representation, with the same charge and statistics but
different spin polarization. The total spin of these states is
$S=np+{\frac{1}{2}}$. Thus, these states are not the ``elementary"
electron which must be a fermion with $Q=1$ and $S={\frac{1}{2}}$.

Actually, a set of states of $N$ elementary quasiparticles, each
with $S={\frac{1}{2}}$, spans a Hilbert space of dimension $2^N$.
Clearly, the states produced by the construction indicated above
seems to give fewer states, apparently only the states with the
largest total spin. In order to construct all the states created
by a set of sources labeled by the pairs of integers $(k_1,k_2)$,
it is necessary to compute the wave functions of the effective
Chern-Simons theory in the presence of the sources labeled by
$(k_1,k_2)$. It is well known from Chern-Simons
theory\cite{witten,israelies,jackiw} that, in order to define the
states created by a set of sources, a set of observables which act
as canonically conjugate pairs need to be chosen. In Chern-Simons
theory this is  called a choice of polarization. For a
Chern-Simons theory on a disk, a standard choice is to select
$A_z=A_1+iA_2$ as the ``coordinate" and $A_{{\bar z}}$ as the
conjugate momentum. This choice is called {\sl holomorphic
polarization}. Within the holomorphic polarization, the wave
functions are holomorphic functions with a singularity structure
determined by the location and quantum numbers of the sources.It
is well known \cite{witten} that these wave functions  coincide
with the (holomorphic) conformal blocks of the correlation
functions of a chiral conformal field theory in two-dimensional
Euclidean space. This correspondence is equivalent to the
statement that there is a one-to-one correspondence between the
states of the bulk and the correlation functions at the boundary.

For the bilayer problem that we are interested in here, these
considerations imply that two sets of complex variables, $z$ and
$w$, one for each layer, are needed to label the states. The full
$2^N$-dimensional Hilbert space of $N$ quasiparticles is obtained
by proper symmetrization and antisymmetrization of the (product)
wave functions, as required by the $SU(2)$ symmetry.  Since this
construction of the bulk wave functions of the excitations
coincides with the correlation functions at the edge, we will
discuss it in detail in the next sections where the correlation
functions at the edge will be constructed. In what follows will
refer to the ``elementary " electron as the state with unit
charge, Fermi statistics and spin polarization $\pm
{\frac{1}{2}}$. The unique choice that satisfies all of these
requirements is the properly antisymmetrized state
 obtained from a set of sources with quantum numbers $(k_1,k_2)=((m-1)p+1,(m-1)p)$
for an electron with spin up, and $(k_1,k_2)=((m-1)p,(m-1)p+1)$
for an electron with spin down respectively. This choice also
works for all $U(1) \times U(1)$ symmetric states.

Finally, the excitations of polarized $SU(2)$ states, {\it i.\
e.\/} states with $S^z_{\rm total} \neq 0$, can be constructed in
a similar manner. The quantum numbers of their excitations are the
same as in the unpolarized case except that the polarization $S_z$
is shifted downwards by $- {\frac{n \nu}{1-n\nu}} Q_{qp} S^z_{\rm
total}$. Hence, these states also form representations of $SU(2)$.
The only change is that the spin projections of the excitations
are shifted by a constant amount determined by the filling
fraction, the charge of the excitation and the polarization of the
ground state.

Let us look in particular at the spectrum of quasiholes and
electrons for the spin singlet state $(3,3,2)$ with filling
fraction $\nu={\frac{2}{5}}$. This state  has $p_1=p_2=1$ and
$n=2$. Notice that, whether or not all of the states discussed
below actually occur in the spectrum of a 2DEG with a specific
Hamiltonian depends on the details of the interactions. In
general, if the Hamiltonian is invariant under the global $SU(2)$
rotations of spin there will not be any matrix elements that will
mix these states. However, even if the $SU(2)$ symmetry were
exact, in general {\it bound states} with all of these quantum
numbers will not be present. Naturally, the excitation with lowest
charge (the quasiholes) is realized.
\begin{enumerate}
\item
Elementary Quasiholes:\\ The elementary quasiholes are vortices
which have $(k_1,k_2)=(1,0), (0,1)$. Their charge is $Q=1/5$, and
there are two polarization states with $S_z=\pm 1/2$. The
statistics of the quasiholes is
${\frac{\theta}{\pi}}=1+{\frac{2}{5}}$. This excitation is always
present in the spectrum.
\item
Neutral Quasihole Bound States:\\ The neutral bosonic bound states
are the collective modes of the fluid. They are described by the
fluctuations of the Chern-Simons gauge fields. We already
discussed their spectrum in reference \cite{paper2}. From the
point of view of the construction that we are discussing here, the
neutral $Q=0$ collective (bosonic) modes have $k_1+k_2=0$. There
is a spectrum of in-phase (charge fluctuations) and out-of-phase
(spin fluctuations) collective modes.
\item
Charged Quasihole Bound States:\\ The simplest charged quasihole
states have $(k_1,k_2)=(2,0),(1,1),(0,2)$. Their charge is
$Q=2/5$, the spin polarization is $S_z=0,\pm 1$, and their
statistics is ${\frac{\theta}{\pi}}=5+{\frac{3}{5}}$. This is the
$S=1$ spin triplet representation of $SU(2)$. It is obvious that
there   should be an $SU(2)$ spin {\sl singlet} as well. This
state is obtained by antisymmetrization of the two-quasihole
states. We will construct this state  in section \ref{sec:op} when
we discuss the realization of these states at the edge.
\item
Electron States:\\ The electron states are bound states of five
elementary quasiholes. Thus, they are created by sources with
quantum numbers $(5,0), (4,1),(3,2),(2,3), (1,4),(0,5)$. These
states have charge $Q=1$. A naively constructed bound state is
fully symmetric and it forms an $SU(2)$ multiplet whose spin
polarizations are $S_z={\pm \frac{5}{2}},{\pm \frac{3}{2}},{\pm
\frac{1}{2}}$. These electron states are fermions since their
statistics is ${\frac{\theta}{\pi}}=35$. Hence, it can be viewed
as a spin $5/2$ multiplet constructed with five quasiparticles of
both types. This multiplet is thus a set of states with maximal
total spin. Obviously, there are other multiplets of five
$S_z=1/2$ charge $1/5$ quasiparticles, forming fermionic bound
states with charge $1$ and with total spin $3/2$ and $1/2$. In
these states, pairs of quasiparticles are placed in spin singlets.
In particular, the actual electron state, with $S=1/2$, is a
product of two singlets and one doublet of quasiparticle states.
These features are generally present in all the $SU(2)$-symmetric
states. These issues will play an important role in Section
\ref{sec:op} where we give details of the construction of the
electron operator at the edge.
\end{enumerate}

It is also instructive to look at the spectrum of excitations of a
partially polarized $\nu=2/5$ state. Now we choose $p_1\neq p_2$
but keep $p_1+p_2=2$ and $n=2$. The spectrum is still classified
formally by the representations of $SU(2)$. For this particular
filling fraction the allowed values of $p_1$ and $p_2$ are $0$ and
$2$. Hence, these states are maximally polarized. As far as the
quantum numbers are concerned, the effect of the asymmetry is a
shift in the value of $S_z$ by an amount determined by the total
polarization of the ground state, the filling factor and the
charge of the state. Of course, this shift is not consistent with
$SU(2)$. Thus, these states have fractional z-projection of the
spin. Notice that, although there are no representations of
$SU(2)$ with fractional quantum numbers ({\it i.\ e.\/} not
integers or half-integers), there are such representations for the
$U(1)$ subgroup of $SU(2)$ generated by $S_z$ since this group is
Abelian. Hence, once the $SU(2)$ symmetry is broken, fractional
quantum numbers for z-component of the spin are allowed. This is
the case for generic $U(1) \times U(1)$ FQH states. For instance
for the quasihole $(k_1,k_2)=(1,0)$ ($(k_1,k_2)=(0,1)$), the spin
polarization is $S_z=1/10$ ($S_z=-9/10$) given that $S_z^{\rm
total}=1/2$. In the presence of a non-zero Zeeman interaction the
energy of these two quasiholes states are not the same, and the
lowest energy quasiholes has $S_z=1/10$.

States with the quantum numbers of electrons can be constructed in
the usual way, {\it i.\ e.\/} by binding $5$ quasiparticles.
However, here the electron with maximal $S_z$ is simply the tensor
product of five quasiparticles with $S_z=1/2$, and it coincides
with the usual electron of a fully polarized $\nu=2/5$ Jain state.

Another state of interest is the singlet state at $\nu=2/3$ which
has been observed experimentally\cite{singlet-23}. This state has
$p_1=p_2=-1$ and $m=n+1=3$. Thus it belongs to the reversed sequence of
the $(3,3,2)$ FQH state. Transitions between the singlet state at
$\nu=2/3$ and the fully polarized state at the same filling factor
have also been observed\cite{singlet-23}. The excitation spectrum
of the $\nu=2/3$ spin singlet state can be constructed
analogously.

\subsection{$U(1) \times U(1)$ FQH states}
\label{subsec:general}

Earlier in this section we gave formulas that describe the
excitation spectrum of generic $U(1) \times U(1)$ FQH states. Here
we will discuss the excitation spectra of symmetric $U(1) \times
U(1)$ FQH states.

These states are given by the choices $2n_1=2n_2=2s=m-1$ and $n
\neq m-1$. For simplicity we will restrict our discussion to the
symmetric states $p_1=p_2=p$. The special case of the primary
states at $p=1$ are the Halperin states $(m,m,n)$ where $m=2s+1$.
In an earlier publication\cite{paper2} we presented a theory of
these states, including the spectrum of collective modes for the
$\nu=1/2$ $(3,3,1)$ FQH state. The filling fraction of these
states is
\begin{equation}
\nu={\frac{2p}{(m+n-1)p+1}} \label{eq:nummn}
\end{equation}
By construction, these ground states have zero polarization
$S_z=0$, {\it i.\ e.\/} both layers are equally populated.

A generic quasiparticle state, with quasiparticle numbers
$(k_1,k_2)$, has charge $Q$,
\begin{equation}
Q={\frac{k_1+k_2}{(m+n-1)p+1}} \label{eq:Qmmn}
\end{equation}
and polarization $S_z$,
\begin{equation}
S_z={\frac{k}{2\left((m-n-1)p+1\right)}} \label{eq:Mmmn}
\end{equation}
where we have introduced the integer $k=k_1-k_2$. The excitations
can be classified by their charge $Q$ and by the integer $k$. The
statistics of these states is
\begin{eqnarray}
{\frac{\theta}{\pi}}=&&Q^2 \left((m+n-1)p+1\right)^2 \nonumber \\
&&+ {\frac{Q^2}{2}}  \left((m+n-1)p+1\right)\left(m+n-1\right)
\nonumber \\ &&+{\frac{k^2}{2}} {\frac{(m-n-1)}{(m-n-1)p+1}}
\nonumber \\ && \label{eq:statmmn}
\end{eqnarray}
In particular, we must choose $k_1+k_2=1$ for the quasihole states
whose charge $Q$, polarization $S_z$ and statistics $\theta$ are
\begin{eqnarray}
Q=&&{\frac{1}{(m+n-1)p+1}} \nonumber \\
S_z=&&{\frac{k}{2\left((m-n-1)p+1\right)}} \nonumber \\
{\frac{\theta}{\pi}}=&&1+{\frac{m+n-1}{2\left((m+n-1)p+1\right)}}+
{\frac{m-n-1}{2\left((m-n-1)p+1\right)}}  \nonumber
\\
\label{eq:qnsmmn}
\end{eqnarray}
The electron states must have charge $Q=1$, hence for these states
we require $k_1+k_2=(2s+n)p+1$. However, the polarization $S_z$,
is still given by Eq.\ \ref{eq:qnsmmn}. The statistics of the
$Q=1$ states is given by
\begin{eqnarray}
{\frac{\theta}{\pi}}=&&\left((m+n-1)p+1\right)^2 +{\frac{k^2}{2}}
{\frac{\left(m-n-1\right)}{
\left(\left(m-n-1\right)p+1\right)}}\nonumber
\\ &&+ {\frac{1}{2}}  \left((m+n-1)p+1\right)\left(m+n-1\right)
 \label{eq:statQmmn}
\end{eqnarray}
In general these states are not fermions. However, the states with
$k_1+k_2=(m+n-1)p+1$ and $|k_1-k_2|=(m-n-1)p+1$ have charge $Q=1$
and $S_z=\pm1/2$. These composite objects are the electron
excitations of the $U(1)\times U(1)$ FQH states on bilayers.

Let us now discuss the special case of the $(3,3,1)$ FQH state,
with filling factor $\nu=1/2$. Many of the properties of the
excitation spectrum of more general states are already present in
the case of the $(3,3,1)$ state. This state has $2s=2$, $n=1$ and
$p=1$, and $\nu=1/2$.
\begin{enumerate}
\item
Charged Quasiparticles:\\ The excitation with smallest charge is a
particle-hole doublet with charge $\pm 1/4$ and polarization
$S_z=\pm 1/4$ ($Q^s=\pm 1/2$). Since in general there is no
particle-hole symmetry, the apparent degeneracy of this doublet is
always lifted by terms not present in this effective theory. The
statistics of the elementary quasiparticles is $\theta/\pi=1+5/8$.
Likewise there are other quasiparticle bound states (also
doublets) with charges $Q=\pm 1/2, \pm 3/4$, spin polarizations
$S_z=0, \pm 1/2$ and $S_z=\pm 3/4, \pm 1/4$ respectively. Their
statistics are $\theta/\pi=5+1/2$ (for $Q=\pm 1/2$, $S_z=0$), and
$\theta/\pi=6+1/2$ (for $Q=\pm 1/2$, $S_z=\pm 1/2$),
$\theta/\pi=14+5/8$ (for $Q= \pm 3/4$, $S_z=\pm 3/4$), and
$\theta/\pi=12+5/8$ (for $Q= \pm 3/4$, $S_z=\pm 1/4$).
\item
Neutral Fermions:\\ An interesting feature of the primary ($p=1$)
$(3,3,1)$ state is the existence of {\sl neutral fermionic}
excitations\cite{milovanovic}. These states have zero charge,
$Q=0$, layer polarization $S_z=\pm 1/2$ ($Q^s=\pm 1$), and
statistics ${\frac{\theta}{\pi}}=1$. Neutral bound states exist
for all $U(1) \times U(1)$ FQH bilayer states with $m-n$ even and
$p$ odd, {\sl e.\ g.\/} the state $(3,3,1)$ itself and its $p$ odd
descendants. These excitations have $S_z=\pm 1/2$ ($Q^s=\pm 1$),
and $Q=0$. These bound states are {\sl bosons} if $m-n=4s$, and
fermions if $m-n=2s$ (with $s$ odd). The sources that create the
neutral spin $1/2$ excitations have the form $(k_1,k_2)=\pm
{\frac{k}{2}} (1,-1)$, where $k=p(m-n-1)+1$ must be even.
\item
Neutral Bosons:\\ Similarly, the spectrum has neutral bosons with
{\sl integer} $S_z$. These bosonic states are just the collective
modes of the bilayer system, {\it i.\ e.\/} excitations of the
out-of-phase Chern-Simons gauge field (see reference
\cite{paper2}).
\item
Electron States:\\ The electron states have charge $Q=-1$ and are
bound states of four elementary quasiparticles. Likewise, there
are also hole states with charge $Q=1$, which are bound states of
four quasiholes. In the $(3,3,1)$ both sets form particle-hole
doublets (naturally, this degeneracy is also lifted). The states
with $Q=1$ are created by sources with the quantum numbers
$(4,0),(3,1),(2,2),(1,3),(0,4)$. Of these five states with charge
$1$ only two, the states denoted by $(3,1)$ and $(1,3)$, {\it i.\
e.\/} with polarization $S_z=\pm 1/2$, are fermions with
$\theta/\pi=11$, and represent the actual electron states. The
three remaining states have $\theta/\pi=10$ (for
$(k_1,k_2)=(2,2)$), $\theta/\pi=14$ (for $(k_1,k_2)=(4,0),(0,4)$),
and are bosons with charge $1$. The states with $Q=-1$ are
constructed by reversing the sign of $(k_1,k_2)$.
\end{enumerate}
As far as the bulk states are concerned it is obvious that, for a
2DEG with general pair interactions, many of these apparently
degenerate states will not have the same energy, even if
particle-hole symmetry was exact. However, at the edge the
situation may be different. In fact, except for possible
differences in the effective velocities of the edge modes, the
spectrum may be degenerate, not only at the level of the
dimensions of the operators but also in terms of their energies.

Recently it has become clear that the properties of the elementary
quasiparticles (or vortices) of the $(3,3,1)$ FQH state can be
related to the existence of underlying pairing correlations
present in this state. In fact, it has been shown that the
$(3,3,1)$ Halperin state can be regarded as an Abelian paired Hall
state\cite{bonesteel}. It has also been shown that the $(3,3,1)$
Halperin state is closely related to the  non-Abelian Pfaffian
paired Hall state\cite{chetan2,read-rezayi}.

\subsection{The $(m,m,m)$ states.}
\label{sec:mmm}

In this section we have derived an effective low energy theory for
a generic FQH state in a bilayer system, and given an explicit
form for the effective Lagrangian. However, in the case of the
Halperin $(m,m,n)$ states the effective low energy theory is
considerably simpler. These states have been described in detail
in the literarture\cite{wenzee-matrix}, particularly by
Milovanovic and Read\cite{milovanovic} who gave a detailed
description of their excitation spectra using a standard hierarchy
construction. In the construction that we are presenting here (see
also ref.\ \ref{ref:paper2}) the $(m,m,n)$ states have level
$p_1=p_2=1$. It is easy to see, by direct inspection of their
effective $K$ matrix Eq.\ \ref{eq:Keff}, that for all of these
states the fields $a_\mu^\alpha$ and $e_\mu$ do not contribute to
the quantum numbers of the excitations. Hence these fields are
redundant and can be integrated out. The effective Lagrangian of
these states thus involves only  the two fields $b_\mu^\alpha$,
and it has the much simpler form
\begin{eqnarray}
{\cal L}_{\rm eff}[b] =
 - {\frac{2{\tilde T}^{\alpha,\beta}}{4 \pi}}\epsilon_{\mu\nu\lambda}
b_{\mu}^\alpha \partial_{\nu} b_{\lambda}^\beta &-& {\frac{1}{2
\pi}} \epsilon_{\mu\nu\lambda} A_{\mu}^\alpha
\partial_{\nu} b_{\lambda}^\alpha
\nonumber \\ + b_{\mu}^\alpha j_{qp,a}^{\mu} &+& {\cal L}_{\rm
Maxwell}[b] \nonumber \\ && \label{eq;eff-mmn}
\end{eqnarray}
where
\begin{equation}
2 {\tilde T}_{\alpha \beta}= \left(
\begin{array}{cc}
2n_1+1        & {n}\\ n & 2n_2+1
\end{array}
\right) \equiv \left(
\begin{array}{cc}
m_1        & {n}\\ n & m_2
\end{array}
\right) \label{eq:T-tilde}
\end{equation}
Here ${\cal L}_{\rm Maxwell}[b]$ represents subleading
Maxwell-like terms, which are important for the case of the
$(m,m,m)$ states for which the matrix ${\tilde T}$ is degenerate.
In the basis
\begin{eqnarray}
b_\mu^+&=&{\frac{1}{\sqrt{2}}}(b_\mu^1+b_\mu^2) \nonumber \\
b_\mu^-&=&{\frac{1}{\sqrt{2}}}(-b_\mu^1+b_\mu^2) \nonumber \\ &&
\label{eq:basis1}
\end{eqnarray}
the effective Lagrangian for the $(m,m,n)$ states separates into
the Lagrangians for the two (independent) fields $b_\mu^+$ and
$b_\mu^-$, with Chern-Simons coupling constants $m+n$ and $m-n$
respectively\cite{milovanovic,wen},
\begin{eqnarray}
{\cal L}_{\rm eff}[b] &=&- {\frac{m+n}{4 \pi}}
\epsilon_{\mu\nu\lambda} b_\mu^+ \partial_\nu b_\lambda^+
-
{\frac{m-n}{4 \pi}} \epsilon_{\mu\nu\lambda} b_\mu^- \partial_\nu
b_\lambda^- \nonumber \\ &&-{\frac{1}{2 \pi}}
\epsilon_{\mu\nu\lambda} A_{\mu}^\pm
\partial_{\nu} b_{\lambda}^\pm + b_{\mu}^\pm
j_{qp}^{\pm} + {\cal L}_{\rm Maxwell}[b] \nonumber \\ &&
\label{eq:Leff-mmn}
\end{eqnarray}
Milovanovic and Read have given an extensive description of the
excitations of the primary (level $p_1=p_2=1$) Halperin states
using this form of the effective Lagrangian. Naturally their
results for these Halperin states are the same ones we discussed
in the previous subsections.

For the special case of the $(m,m,m)$ states it is also convenient
to introduce the basis
\begin{eqnarray}
b_\mu^+&=&b_\mu^1+b_\mu^2 \nonumber \\
b_\mu^-&=&{\frac{1}{2}}(-b_\mu^1+b_\mu^2) \nonumber \\ &&
\label{eq:basis2}
\end{eqnarray}
The effective Lagrangian for the $(m,m,m)$ states becomes
\begin{eqnarray}
{\cal L}_{\rm eff}[b] &=&-{\frac{m}{4
\pi}}\epsilon_{\mu\nu\lambda} b_{\mu}^+ \partial_{\nu}
b_{\lambda}^+ -{\frac{1}{2 \pi}} \epsilon_{\mu\nu\lambda}
A_{\mu}^+
\partial_{\nu} b_{\lambda}^+ + b_{\mu}^+
j^\mu_{qp,+} \nonumber \\ && +{\frac{1}{2g}}
\left({\frac{1}{v}}{\vec {\cal E}}_-^2-v {\cal B}_-^2\right)
-{\frac{1}{2 \pi}} \epsilon_{\mu\nu\lambda} A_{\mu}^-
\partial_{\nu} b_{\lambda}^- + b_{\mu}^- j^\mu_{qp,-}
\nonumber \\ && \label{eq:Leff-mmm}
\end{eqnarray}
where
\begin{equation}
\begin{array}{ccc}
A^\mu_+     ={\frac{1}{2}}(A^\mu_1+A^\mu_2) & , \;\; &
j^\mu_{qp,+}={\frac{1}{2}}(j^\mu_1+j^\mu_2)\\ A^\mu_-=
-A^\mu_1+A^\mu_2  & , \;\; & j^\mu_{qp,-}=
-j^\mu_1+j^\mu_2
\end{array}
\end{equation}
In the effective Maxwell term of Eq.\ \ref{eq:Leff-mmm}, the
approximate value of the effective coupling constant is $g=2\pi
\sqrt{{\frac {B \nu} {2}}}$, and the approximate effective
velocity is $v=\sqrt{{\frac{\nu B}{2 M^2}}}$, $\nu=1/m$ is the
filling factor of the $(m,m,m)$ state, and $M$ is the electron
mass (see ref.\ [\ref{ref:paper2}]). Eq.\ \ref{eq:Leff-mmm} shows
that the finite energy excitations of the charge sector of the
$(m,m,m)$ state coincides with the charge sector of the
corresponding single layer Laughlin state, and that the neutral
excitations include a gapless mode, the ``photon" of the effective
Maxwell theory.

Wen and Zee\cite{wenzee-gapless} where the first to note that in
this state there is a gapless mode with speed $v$. This mode has
been identified with the Goldstone boson for the ``spontaneous
development of interlayer coherence"\cite{ezawa,indiana}. The
order parameter of this condensate can be described in terms of
the dual picture. In $2+1$ dimensions a gauge theory is always
dual to a theory of a scalar field. We define a real antisymmetric
tensor field  $\Gamma_{\mu\nu}$ such that $\Gamma^{0i}=e_-^i$ and
$\Gamma_{ij}= \epsilon_{ij} b_-$, with $e_-^i$ and $b_-$ real
functions of space and time. It is simple to check that given the
following Lagrangian
\begin{equation}
{\cal L}´ =-{\frac{g}{2}} \left({v}{\vec { e}}_-^{\;2}- {\frac {1}{
v}} { b}_-^{\; 2}\right) -{\frac{1}{2 }} \Gamma^{\mu\nu} f^{-}_{\mu\nu}
\label{eq:Ldual-mmm}
\end{equation}
upon the integration of $\Gamma_{\mu\nu}$, one obtains the Maxwell
term of Eq. \ref{eq:Leff-mmm} for $b_{\mu}^-$, except for an
irrelevant constant.

The integral over $b_{\mu}^-$ in Eq. \ref{eq:Ldual-mmm} gives the
constraint $\partial_{\mu}\Gamma^{\mu\nu}=0$. This constraint can
be solved by means of the phase field $\theta$ by
$\Gamma^{\mu\nu}= {\frac {1}{2\pi}} \epsilon_{\mu\nu\lambda}
\partial^\lambda \theta$ . By substituting back into the effective
Lagrangian, we get the effective action for the  field $\theta$,
dual to the gauge field $b_\mu^-$
\begin{equation}
{\cal L}[\theta]={\frac{{ g}}{8\pi^2}}
\left({\frac{1}{v}}(\partial_0 \theta+2A^0_-)^2-v({\vec
{\nabla}\theta+2{\vec A}_-} )^2\right) \label{eq:Ltheta}
\end{equation}

Notice that in the presence of the sources (currents)
$j_{qp,-}^\mu$ the field $\theta$ is multivalued. Conversely, the
operator $\exp(i\theta)$ creates a monopole configuration of the
field $b_\mu^-$. This is natural since in $2+1$ dimensions the
dual of the order parameter field $\exp(i\theta)$ is a monopole
creation operator \cite{FS}.

Eq.\ \ref{eq:Ltheta} has a number of important features: (i) as
expected, duality has replaced the coupling constant by its
reciprocal divided by  $(2\pi)^2$ , and (ii) the field $\theta$
behaves like the phase field of an order parameter with neutral
charge $2$, a feature emphasized by Ezawa and
coworkers\cite{ezawa}. This form of the dual theory, Eq.\
\ref{eq:Ltheta}, is familiar from the theories of anyon
superfluidity (see for example ref.\ [\ref{ref:anyon}]).

The correlation function of the order parameter is
\begin{equation}
\langle e^{\displaystyle{i\theta(x)}} \;
e^{\displaystyle{-i\theta(x')}} \rangle
=e^{\displaystyle{\frac{\pi}{ g} \left(\frac{1}{ R} -\frac{1}{a}
\right) }} \label{eq:opcr}
\end{equation}
where (in imaginary time $x_0$)
\begin{equation}
R^2=v^2 |x_0-x_0'|^2+|{\vec x}-{\vec x}\;'|^2+a^2
\end{equation}
where $a$ is a short distance cutoff.

At equal times the correlation function is (for large distances
compared to the cyclotron length)
\begin{equation}
\langle e^{\displaystyle{i\theta(0,{\vec x})}} \;
e^{\displaystyle{-i\theta(0,{\vec x}\;')}} \rangle \approx
e^{\displaystyle{\frac{\ell_0}{\sqrt{2\nu} |{\vec x}-{\vec
x}\;'|}}} \label{eq:opcret}
\end{equation}
where $\ell_0=1/\sqrt{B}$ is the cyclotron length. This equal-time
correlation function has the same form as the contribution of the
neutral sector to the wave function of the $(m,m,m)$ state found
in ref.\ [ \ref{ref:paper2}].

One may also suspect that there may be more excitations than the
ones we discussed so far. For example, one possibility is a
quasiparticle (or quasihole) of a single layer. However, these
states carry a non-zero polarization (or ``neutral charge"), and
as such they couple to the Maxwell gauge field $b_\mu^-$. In
addition, precisely because there is a massless mode, the self
energy of quasiparticles that carry a non-vanishing polarization
(neutral charge) is  logarithmically divergent. Such excitations
are thus confined to neutral pairs with respect to the neutral
charge. These ``composite fermions" are actually the vortices of
the condensed state\cite{capitulo,indiana}.

Finally it is useful to write the dual form of the Wilson loop
operators for the gauge field $b_\mu^-$. Consider a closed loop
$\Gamma$ and a quasiparticle current with charge $k_-=k_1-k_2$.
The dual of this Wilson loop operator is
\begin{equation}
\langle e^{\displaystyle{i k_-\oint_\Gamma dx_\mu
b^\mu_-}}\rangle= \langle e^{\displaystyle{i 2\pi k_- \int_\Sigma
dS_\mu \partial^\mu \theta}} \rangle \label{eq:WLb-}
\end{equation}
where $\Sigma$ is an open surface whose boundary is $\Gamma$.

Let us summarize the excitation spectrum of the bulk $(m,m,m)$
states:
\begin{enumerate}
\item
There are electrically charged $e/(2m)$ vortices with
logarithmically divergent energy, statistics $\pi/(4m)$, and
``neutral charge" $\pm 1$, which simply says that there is one for
each layer. These excitations are the ``composite fermions" of the
individual layers.
\item
There are electrically charged $e/m$ finite energy excitations and
statistics $\pi/m$; these excitations are vortex-antivortex bound
states, and are the analogs of the Laughlin quasiholes of a single
layer system.
\item
There are two types of electron states:
\begin{enumerate}
\item
a state with electric charge $-e$, zero polarization and finite
energy. This is the electron state of the Laughlin sector. It is a
bound state with  $m$  negatively charged vortices  of one layer
and $m$ positively charged vortices of the other layer. Here we
refer to the charge coupled to the field $b^-_\mu$.
\item
two states with electric charge $-e$, polarization $\pm 1$ and
logarithmically divergent energy; these states are electrons made
entirely from $2m$ negatively charged vortices all in the same
layer.
\end{enumerate}
\item
There is also a gapless mode or Goldstone boson.
\end{enumerate}
In section \ref{sec:mmm-edge} we will discuss the spectrum of edge
states.

\section{Edge theory for FQH states on bilayers and partially polarized states}
\label{sec:edge}

In this section we use the  theory derived in the previous
section, to extract the effective theory for the edge states. The
effective Lagrangian of Eq.\ (\ref{eq:eff9}) is globally well
defined (on closed surfaces), and has excitations with the correct
fractional charge and statistics. This effective Lagrangian has
the standard form first introduced by Wen and
Zee\cite{wenzee-matrix}. Following the general arguments of
Wen\cite{wen,wen-edge,stone-edge}, it is straightforward to
extract a theory for the edge states which reflects the structure
of the bulk presented in the previous section.

 The Hilbert space of states of a Chern-Simons theory on a
manifold $\Omega$ (which we take to be a disk) with a spatial
boundary $\partial \Omega \cong S_1$ (where $S_1$ is a circle) has
support at the boundary. This is a special case of a general
result originally derived by Witten\cite{witten}. We now imagine
that there is a (sharp) potential that confines the electrons on
some simply connected region on the torus, isomorphic to a disk.
The gauge fields on the region forbidden to the electrons can be
integrated out since they decouple. Furthermore, we will {\rm
assume} that the confining potential is sharp enough that there is
no edge reconstruction. Whether or not this assumption is
justified depends on (important) microscopic details, including
the form of the interactions, the vicinity of gates, etc.\ It is
well known\cite{chklovskii} that the mean field theory
approximation to the fermion Chern-Simons approach to the FQHE
predicts a rather delicate structure of the edge modes. However,
provided the assumption that the edge is sharp holds, we can
assume that the spatial separation of these edge branches is
small, \ie\ of the order of the magnetic length. In this case,
even if these modes were to exist, something that in this regime is highly
questionable, it is clear that they cannot be resolved as separate Hilbert spaces. Furthermore, these modes will
couple to the same gauge field(s). Thus, we will adopt the physical point of view that, in this regime, it is
legitimate to ``glue" these modes together and to treat them as if they were a single mode with an effective
velocity. Notice that, under these assumtions and due to the renormalization of the mode velocity found below, 
the contribution of $p$ edge modes glued together to the specific heat
is the same as the contribution of $p$ independent modes. Thus, this assumption does not change the physics.
Moreover, below we will find that the structure of the bulk theory on a torus requires the existence of a set of
non-propagating modes at the boundary (which we will refer to as {\sl topological modes}). These topological modes
are {\sl unrelated} to a possible fine structure of the edge and play a very different role. In particular they do
not carry energy and only show up in the statistics if the physical states\cite{ref1}.

Thus, in this regime we can derive an effective theory of the edge states
directly from the effective action of  Eq.\ (\ref{eq:eff9}) in its present form without further assumptions. It is
straightforward to show that the effective action can be expanded
in the form
\begin{eqnarray}
S=&&{\frac{1}{4\pi}} K_{IJ} \int_\Omega d^3x \;\;
\partial^j \left(a_J^0\epsilon_{ij}a_I^i\right)
\nonumber\\ +&&{\frac{1}{4\pi}} K_{IJ} \int_\Omega d^3x \;\; a_I
^0 \epsilon_{ij} \left(\partial^i a_J^j-\partial^j a_I^i\right)
\nonumber\\ -&& {\frac{1}{4\pi}} K_{IJ} \int_\Omega d^3x \;\;
\epsilon_{ij} a_I^i \partial^0  a_J^j \nonumber\\
+&&{\frac{1}{2\pi}}  \int_\Omega d^3x \;\; \left(a_0^I {\cal
J}^0_I+a_i^I {\cal J}^i_I\right) \nonumber\\ &&
\label{eq:effaction1}
\end{eqnarray}
where we have used the current ${\cal J}^i_I$ defined by
\begin{equation}
{\cal J}^\mu_I \equiv t_{aI} \epsilon^{\mu \nu \lambda}
\partial_\nu A_\lambda^a+2\pi \ell_{aI} j^{\mu,a}_{qp}
\label{eq:current}
\end{equation}
We will impose the gauge condition $a_J^0=0$ at the boundary
$\partial \Omega$. In this gauge the first term of Eq.\
(\ref{eq:effaction1}) vanishes. In this form of the action it is
also apparent that the field $a_J^0$  is a Lagrange multiplier
that enforces the local constraint
\begin{equation}
{\cal J}^0_I=-K_{IJ} \epsilon_{ij} \partial^i a_J^j
\label{eq:gauss}
\end{equation}
which is just Gauss' Law. Similarly, the third term of Eq.\
(\ref{eq:effaction1}) determines the commutation relations.

The solution of Gauss' Law is
\begin{equation}
a_i^I=\partial_i \phi^I \label{eq:gauss-solution}
\end{equation}
where $\phi^I$ are five multivalued scalar fields, {\it i.\ e.\/}
singular gauge transformations. If the quasiparticles and the
external fluxes are quasistatic bulk perturbations of the
condensate, of quasiparticle number $N_{qp}^{i}$ and flux
$\Phi^a=2\pi N_\phi^a$ with $a=1,2$, the scalar fields $\phi^I$ at
the boundary $\partial \Omega$ must satisfy the conditions
\begin{eqnarray}
\Delta \phi_I= 2\pi \left(K^{-1}\right)_{IJ} \left[
\begin{array}{c}
N_\phi^{(1)} \\ N_{qp}^{(1)} \\N_\phi^{(2)} \\ N_{qp}^{(2)} \\
-N_{qp}^{(1)}-N_{qp}^{(2)}
\end{array}
\right]_J \label{eq:sources}
\end{eqnarray}
where $\Delta \phi_I\equiv \oint_{\partial \Omega}dx_i
\partial_i\phi_I$ is the change of the field $\phi_I$ once taken around the
boundary $\partial \Omega$.

Once the constraint Eq.\ (\ref{eq:gauss}) is solved, it is
immediate to show that the content of this theory resides at the
boundary $\partial \Omega$. Indeed, the remaining term in the
action Eq.\ (\ref{eq:effaction1}) takes the form
\begin{eqnarray}
S=&&- {\frac{1}{4\pi}} K_{IJ} \int_\Omega d^3x \;\; \epsilon_{ij}
a_I^i \partial^0  a_J^j \nonumber\\ =&&{-\frac{1}{4\pi}}
K_{IJ}\int dx_0 \oint_{\partial\Omega}dx_i
\partial^i\phi_I \partial^0\phi_J
\nonumber\\ && \label{eq:ccr}
\end{eqnarray}
which is  a theory of chiral bosons. However, as emphasized by
Wen\cite{wen}, as they stand these bosons do not propagate. The
reason is that the Chern-Simons gauge theory is  actually a
topological field theory. Thus, in addition to being gauge
invariant, it is independent of the metric of the manifold where
the electrons reside and hence it is also invariant under
arbitrary local diffeomorphisms. In particular this means that the
Hamiltonian of the Chern-Simons theory is zero. Naturally, this is
just the statement that this is an effective theory for the
degrees of freedom below the gap of the incompressible fluid.
There are no local degrees of freedom left in this regime. The
degrees of freedom only ``materialize" at the boundary which, in
addition to breaking gauge invariance, also break the topological
invariance. This is physically obvious since the edge states at
the boundary carry energy and their Hamiltonian does not vanish.

There are several possible ways to represent this physics in the
effective theory. The generalization of  our approach in reference
\cite{ref1} to the bilayer system is straightforward. The edge
term in the action due to the presence of a confining potential
for a sharp edge will be  given by
\begin{eqnarray}
S_{\rm edge}=&&-{\frac{p_1v_1}{4\pi}} \int dx_0 \oint_{\partial
\Omega} dx_1 \left(\partial_1 \phi^1\right)^2 \nonumber\\
&&-{\frac{p_2v_2}{4\pi}} \int dx_0 \oint_{\partial \Omega} dx_1
\left(\partial_1 \phi^3\right)^2
 \label{eq:bt2}
\end{eqnarray}
with $v_\alpha=eE_\alpha /B$  the speed of the edge excitations
for layer $a=1,2$, and we have used the gauge condition
$a_0^1=a_0^3=0$ at the boundary. At this point we allow for the
possibility of the two physically separated layers to have
different velocities.

The electron-electron interaction term of the Hamiltonian becomes
\begin{eqnarray}
H_{\rm int}&&=\int_x \int_{x'} {\frac{1}{2}} (\rho_a (x)-{\bar
\rho_a}) V_{ab}(x-x') (\rho_b(x')-{\bar \rho_b}) \nonumber\\
&&\equiv \oint_{\partial\Omega} dx_i dx'_i {\frac{t_I^a t_J^b
}{8\pi^2}}
\partial_i \phi_I(x) V_{ab}(x-x') \partial_i \phi_J(x') \nonumber\\ &&
\label{eq:el-el}
\end{eqnarray}
where $a$ and $b$ label the layers. We have only retained the
boundary contribution since the bulk excitations have a finite
(and for present purposes large) energy gap. Notice that, given
the form of $t_{aI}$ as in Eq.\ \ref{eq:t}, the term of the
Hamiltonian of Eq.\ \ref{eq:el-el} only affects the modes $\phi_1$
and $\phi_3$. Likewise, the interaction with an external potential
with support at the boundary becomes
\begin{eqnarray}
H_{\rm ext}=&& -\int d^2x \left( \rho_a(x)-{\bar \rho}_a\right)
A_0^a(x) \nonumber\\ =&& - \oint_{\partial\Omega} dx_i
{\frac{t_{aI}}{2\pi}}
\partial_i\phi_I(x) A_0^a(x)
\nonumber \\ && \label{eq:sext}
\end{eqnarray}
and it involves only $\phi_1$ and $\phi_3$ for the same reason.

In summary the effective action involves five bosons $\phi_I$
(with $I=1,...5$) and takes the form
\begin{equation}
S={\frac{1}{4\pi}}\int_{\partial \Omega \times {\bf R}} dx_0 dx_1
\left( -K_{IJ}
\partial_1\phi_I \partial_0\phi_J- U_{IJ}\partial_1\phi_I \partial_1\phi_J
\right) \label{eq:edge}
\end{equation}
where $U_{IJ}=t_{aI} t_{bJ}\left((\delta_{a1} v_1 p_1
\delta_{b1}+\delta_{a2} v_2 p_2 \delta_{b2})+ {\frac{1}{2\pi }}
V_{ab}\right)$, and its only effect is to determine the velocity
of the edge modes. Notice that, as it is well known, the actual
velocity of the edge modes is the sum of two terms, one of which
is determined by the interactions. The modes with a non vanishing
velocity are $\phi_{1},\phi_{3}$ which are the only ones that
couple to perturbations due to an external electromagnetic field.
Thus we identify $\phi_{1},\phi_{3}$ as the {\it charge modes}.
The three remaining modes do not propagate. Their effect is to fix
the statistics of the states.

Finally, we need to relate these fields to the edge charge density
for each layer. The local charge and current density $J_\mu^a(x)$
in layer $a$  is given by
\begin{equation}
J_\mu^a(x)={\frac{\delta S}{\delta A^\mu_a}}={\frac{1}{2\pi}}
t_{aI} \epsilon_{\mu \nu \lambda} \partial^\nu a^\lambda_I
\label{eq:curr}
\end{equation}
Following the same steps described in reference \cite{ref1} we can
integrate the bulk currents across the edge to obtain the edge
densities and currents. If the physical width of the edge
$\Lambda$ is infinitesimal relative to the linear size of the
system, we find that the edge charge density and current density
for the edge of layer $a$, in the $x_1$ direction, is given by
\begin{eqnarray}
j_0^1\equiv&&\int_\Lambda dx_2 J_0^1(x)=
{\frac{1}{2\pi}}\partial_1 \phi_1 \nonumber \\
j_0^2\equiv&&\int_\Lambda dx_2 J_0^2(x)=
{\frac{1}{2\pi}}\partial_1 \phi_3 \label{eq:densi}
\end{eqnarray}
It is straightforward to check that if $N_{qp}^\alpha$
quasiparticles  of type $\alpha$ are added to the bulk at constant
magnetic field ($N_\phi=0$), the edge acquires a charge
\begin{eqnarray}
Q_{\rm edge}&=&Q_{\rm edge}^{(1)}+Q_{\rm edge}^{(2)}\nonumber \\
&=&\int dx_1 (j_0^1(x_1)+j_0^2(x_1))=N_{qp}^{\alpha}
{\frac{\nu_{\alpha} }{p_{\alpha} }}
\end{eqnarray}
which is, as expected, equal to the extra charge added to the
bulk.

To conclude this section we remark once again  that the
correspondence between the bulk theory described in section
\ref{sec:bl} and the edge theory described here is only correct
for a sharp edge. In this case we found that the theory of the
edge states of the FQH states in bilayers can be described in
terms of five chiral fields. Two of them are propagating fields,
and  are associated with charge fluctuations. The other three
non-propagating fields are associated with the global topological
consistency of flux-attachment. As we will see in the next section
the only effect of these non-propagating topological modes is to
give the correct statistics to the excitations. We will also find
that it is possible to reduce the number of non-propagating
topological fields to just two.

However, in many experimental situations, the confining potentials
are smooth. This implies that the edge scale is of the order of
many magnetic lengths and the density gradually drops to zero
within this scale. In this case the edge will show the usually
called edge reconstruction in which two or more (possibly)
interacting edge branches are generated. Within our approach, this
effect will appear at the level of the mean field solution for the
bulk action( eq \ref{eq:a3B} ). In the presence of a smooth
confining potential in the sample, the calculation of the
fermionic determinant  has to take into account the presence of
different edges and their interactions for the mean field
solution. We will not deal with this problem here.

\section{ Electron and quasiparticle operators for the edge states on bilayers}
\label{sec:op}

In this Section we will construct the operators for electrons and
quasiparticles of the edge states in bilayer FQH states. As we
have shown in section \ref{sec:bl}, all elementary physical
excitations can be written as combinations of $k_1$ quasiparticles
of type $1$ and $k_2$ quasiparticles of type $2$. Therefore a
generic edge operator that creates excitations can be written as
\begin{equation}
\Psi (x) = e^{\displaystyle{i [k_1 \phi_2 + k_2 \phi_4 - (k_1+k_2)
\phi_5]}} \label{eq:opqp}
\end{equation}
 At this stage, it is convenient to rewrite the effective theory
in a new basis where the charged and topological  fields decouple
completely. We have already identified  $\phi_1$ and  $\phi_3$ as
the charge fields, \ie\ the fields  that couple to an external
electromagnetic field. We further introduce the topological fields
$\phi_{T\alpha},\phi_{T0}$ and define the new basis
\begin{eqnarray}
     \phi_{C1}= &&\phi_1
\nonumber \\
      \phi_{T1} =&& {\rm sign}(p_1) {\frac{1}{\sqrt{|p_1|}}} \phi_1 + {\sqrt{|p_1|}} \phi_2
\nonumber \\
 \phi_{C2}= &&\phi_3 \nonumber \\
      \phi_{T2} =&& {\rm sign}(p_2) {\frac{1}{\sqrt{|p_2|}}} \phi_3 + {\sqrt{|p_2|}} \phi_4
\nonumber \\
     \phi_{T0}=&&\phi_5
\nonumber \\ \label{eq:CN}
\end{eqnarray}

The edge effective Lagrangian  of Eq ~(\ref{eq:edge}) in terms of
the charged and topological fields becomes
\begin{eqnarray}
{\cal L}=&&{\frac{1}{4\pi}} \left(\kappa_{\alpha\beta}
 \partial_1\phi_{C\alpha}  \partial_0\phi_{C\beta}
 -  \partial_1\phi_{C\alpha} v_{\alpha\beta} \partial_1\phi_{C\beta} \right)
\nonumber \\ &&-{\frac{1}{4\pi}}\left(G_{\alpha \beta}
\partial_1\phi_{T\alpha}
\partial_0 \phi_{T\beta} + \partial_1\phi_{T0} \partial_0
\phi_{T0} \right) \nonumber \\ && \label{eq:edgenew}
\end{eqnarray}
where
\begin{eqnarray}
\kappa=\left(
\begin{array}{cc}
 2n_1+{\frac{1}{p_1}} & n \\
  n &2n_2+{\frac{1}{p_2}}
\end{array}
\right) \label{eq:kapa}
\end{eqnarray}
\begin{eqnarray}
G=\left(
\begin{array}{cc}
 {\rm sign}(p_1) & 0 \\
  0 &{\rm sign}(p_2)
\end{array}
\right) \label{eq:G}
\end{eqnarray}
and the velocities are given by
\begin{eqnarray}
v= \left(
\begin{array}{cc}
 V_{11} + 2\pi p_1 v_1 & V_{12} \\
  V_{12} &  V_{22} + 2\pi p_2 v_2
\end{array}
\right)\label{eq:velo}
\end{eqnarray}
Notice that we have allowed for $p_1$ and/or $p_2$ to be either
positive or negative.

We see that only the $\phi_{C\alpha}$ modes propagate. The role of
the three remaining modes is to give the correct statistics to the
quasiparticles. In the new basis of Eq.\ \ref{eq:CN}, the most
general quasiparticle operator of Eq.\ \ref{eq:opqp} can be
written as
\begin{equation}
\Psi (x) = e^{\displaystyle{i \left(a_{C\beta} \phi_{C\beta} +
a_{T\beta} \phi_{T\beta}+a_{T0}\phi_{T0}\right)}} \label{eq:opqp2}
\end{equation}
where
\begin{eqnarray}
     a_{C1}=&& -{\frac{k_1}{p_1}}  \;\;\;\;\;\;\;\;\;\;
     a_{C2}=   -{\frac{k_2}{p_2}}
\nonumber \\
     a_{T1} =&& {\frac{k_1}{\sqrt{|p_1|}}} \;\;\;\;\;\;\;\;\;\;
     a_{T2}=    {\frac{k_2}{\sqrt{|p_2|}}}
\;\;\;\;\;\;
 a_{T0}=-(k_1+k_2)
\nonumber \\ && \label{eq:coeff}
\end{eqnarray}
As expected, the quantum numbers of the states created by these
operators are given by Eq.\ \ref{eq:totQM} and Eq.\
\ref{eq:statcomp}.

For the quasiparticle (and quasihole) operators we choose
$(k_1,k_2)=(1,0)$ for the quasiparticle of type $1$, and
$(k_1,k_2)=(0,1)$ for the quasiparticle of type $2$ respectively,
\begin{eqnarray}
\Psi_{\downarrow} =&& e^{\displaystyle{i \left(-{\frac{1}{p_1}}
\phi_{C1} + {\frac{1}{\sqrt{|p_1|}}} \phi_{T1} -\phi_{T0}\right)}}
 \nonumber \\
 \Psi_{\uparrow} =&& e^{\displaystyle{i
\left(-{\frac{1}{p_2}} \phi_{C2} + {\frac{1}{\sqrt{|p_2|}}}
\phi_{T2} - \phi_{T0}\right)}} \nonumber \\ && \label{eq:opqp3}
\end{eqnarray}
where we have renamed the quasiparticles as $\Psi_{\downarrow}$
for $(k_1,k_2)=(1,0)$, and $\Psi_{\uparrow}$  for
$(k_1,k_2)=(0,1)$, depending on whether their spin projection
$S_z$ is negative or positive respectively.

~From now on we will discuss the special case in which the
Lagrangian \ref{eq:edgenew} takes a simple (diagonal) form. In
particular we will set $2n_1+1 =2n_2+1\equiv m$, and
$p_1=p_2\equiv p$. We will also assume that the velocities in both
layers are equal. Let us define the rotated and rescaled fields
$\phi_{C\pm}$ and $\phi_{T\pm}$,
\begin{eqnarray}
\phi_{C\pm}=&&{\displaystyle{\sqrt{\frac {|m-1+\frac{1}{p}\pm
n|}{2}}}} \left( \phi_{C1} \pm \phi_{C2}\right) \nonumber \\
\phi_{T\pm}=&&{\frac {1}{\sqrt{2}} }\left( \phi_{T1} \pm
\phi_{T2}\right) \nonumber \\ && \label{eq:fipm}
\end{eqnarray}
which will simplify the description considerably. The effective
velocities $v_\pm$ are
\begin{equation}
v_\pm=\displaystyle{\frac{ V_{11} + 2\pi p v \pm
V_{12}}{\sqrt{|m-1+\frac{1}{p}\pm n|}}} \label{eq:vs}
\end{equation}
where we have assumed $v_1=v_2=v$. We will also need the
definition
\begin{equation}
s={\rm sign} \left(m - n-1+\frac{1}{p}\right) \label{eq:sign}
\end{equation}
The sign $s$ of Eq.\ \ref{eq:sign} determines the chirality of the neutral (or spin) edge state, relative to the
chirality of the charge mode.
Recently, Moore and Haldane have found in an exact diagonalization in small systems that in the $2/3$ singlet
state the charge and spin edge modes have opposite chirality\cite{moore-haldane}. 
This result is consistent with the general rule of Eq.\ \ref{eq:sign}.

Furthermore, it is  possible to find still another basis in which
one of the topological non-propagating fields decouples
completely. The details of this procedure depend on the sign of
$p$. For $p > 0$ we rotate the fields $\phi_{T+}$ and $\phi_{T0}$  by
the orthogonal transformation
\begin{eqnarray}
\phi_{T+} \to  && \cos \theta \; \phi_{T+} \; - \sin \theta \;
\phi_{T0} \nonumber\\ \phi_{T0} \to  && \sin \theta \; \phi_{T+}
\; + \cos \theta \; \phi_{T0} \nonumber \\ && \label{eq:OT}
\end{eqnarray}
where the angle $\theta$ is given by
\begin{equation}
\tan \theta=\sqrt{2p} \label{eq:tan}
\end{equation}
whereas for $p<0$ the required transformation has instead the form
of a Lorentz transformation,
\begin{eqnarray}
\phi_{T+} \to && \cosh \theta \; \phi_{T+} \; + \sinh \theta \;
\phi_{T0} \nonumber\\ \phi_{T0} \to && \sinh \theta \; \phi_{T+}
\; + \cosh \theta \; \phi_{T0} \nonumber \\ && \label{eq:HT}
\end{eqnarray}
where the ``rapidity" $\theta$ is now given by
\begin{equation}
\tanh \theta=\frac{1}{\sqrt{2|p|}} \label{eq:tanh}
\end{equation}
Upon a simple redefinition of the fields it is possible to
describe both cases simultaneously. In this condensed notation,
the entire excitation spectrum can be generated in terms of two
propagating chiral bosons, the fields $\phi_{C\pm}$, and two
non-propagating topological fields that we will denote by
$\phi_{T\pm}$. In terms of these fields, the effective Lagrangian
that results is
\begin{eqnarray}
{\cal L}=&&{\frac {1}{4\pi}}
 \partial_1\phi_{C+} (\partial_0\phi_{C+}
      - v_+ \; \partial_1\phi_{C+} )
\nonumber \\
      + &&{\frac {1}{4\pi}}
 \partial_1\phi_{C-} (s \; \partial_0\phi_{C-}
      - v_- \; \partial_1\phi_{C-} )
\nonumber \\ -&&  {\frac  {1}{4\pi}}  \partial_1\phi_{T+}
\partial_0 \phi_{T+} - {\frac{{\rm sign}(p)}{4\pi}} \;  \partial_1\phi_{T-}
\partial_0 \phi_{T-}
\nonumber \\ && \label{eq:edgediag-final}
\end{eqnarray}
The charge and spin currents are given by
\begin{eqnarray}
j_C=&&{\frac{1}{2\pi}}
\sqrt{\displaystyle{\frac{2}{|m+n-1+\frac{1}{p}|}}} \;\;\;\;
\partial_x \phi_{C+}
\nonumber \\ && \label{eq:charge-current}
 \\
j_z=&&{\frac{1}{2\pi}}
\displaystyle{\frac{1}{\sqrt{2|m-n-1+\frac{1}{p}|}}} \;\;\;
\partial_x \phi_{C-}
\nonumber\\ && \label{eq:spin-current}
\end{eqnarray}
The most general edge excitations are created by an operator of
the form
\begin{equation}
\Psi (x) = e^{\displaystyle{i \left(a_{C+} \phi_{C+} + a_{C-}
\phi_{C-} +a_{T+} \phi_{T+}+a_{T-} \phi_{T-} \right)}}
\label{eq:opqp33}
\end{equation}
where, for states with $2n_1 + 1=2n_2+ 1=m$, the coefficients
$a_{C\pm}$ and $a_{T-}$ now take the form
\begin{eqnarray}
     a_{C+}=&& \displaystyle{-{\frac{k_1+k_2}{ p\sqrt{2|m+n-1+\frac{1}{p}|}}} }
     \nonumber \\
     a_{C-}=&& \displaystyle{-{\frac{k_1-k_2}{ p\sqrt{2|m-n-1+\frac{1}{p}|}}} }
     \nonumber \\
     a_{T+} =&&\displaystyle{ {\rm sign}(p) \; \left(k_1 + k_2 \right) \sqrt{1+\frac{1}{2p}}}
    \nonumber \\
     a_{T-} =&&\displaystyle{ {\frac{k_1- k_2}{\sqrt{ 2|p|}}}}
     \nonumber \\
     &&
\label{eq:coeff2}
\end{eqnarray}
Notice that here $p$ can have either sign.

The quantum numbers of the excitations created by the operators of
Eq.\ \ref{eq:coeff2} are
\begin{eqnarray}
     {\frac{Q} {e}}=&&  {\sqrt{\nu}} \; a_{C+}
\nonumber \\
      S_z =&& {\frac{1}{\sqrt{2|m-n-1+\frac{1}{p}|}}} \; a_{C-}
\nonumber \\
        {\frac {\theta}{\pi}}  =&& -  (a_{C+})^2 -s \; (a_{C-})^2
+  (a_{T+})^2 + {\rm sign} (p) \; (a_{T-})^2 \nonumber \\ &&
\label{eq:chst3}
\end{eqnarray}
It is easy to check that the operators
\begin{eqnarray}
\lefteqn{\Psi_{{\rm qp};\uparrow,\downarrow} = e^{\displaystyle{-i
\sqrt{1+ {\frac {1}{2p}}} \; \phi_{T+}\pm {\frac {i}{\sqrt{2|p|}}}
\; \phi_{T-}}}} \nonumber \\ &&  \times \; e^{\displaystyle{
-{\frac {i}{p{\sqrt {2|m+n-1+\frac{1}{p}|}} }} \phi_{C+} \pm
{\frac {i}{p{\sqrt{2|m-n-1+\frac{1}{p}|}} }} \phi_{C-} }}
 \nonumber \\
&& \label{eq:opqp5}
\end{eqnarray}
create excitations with the correct quantum numbers for the
elementary quasiholes.

Next we compute the propagators for an excitation created by an
operator of the form of Eq.\ \ref{eq:opqp33}. Since the effective
action is quadratic in the fields, this calculation is
straightforward. The propagators of the fields $\phi_{C\pm}$ are
identical since their actions are the same. In imaginary time they
become\cite{ref1,cft}
\begin{equation}
\langle \phi_{C\pm}(x,t)  \phi_{C\pm}(0,0) \rangle= -\ln z
\label{eq:gffic}
\end{equation}
where $z=x+iv_{\pm} t$.

For the fields  $\phi_{T\pm}$, which do not propagate ({\it i.\
e.\/} their velocity is zero),  their propagators in the
 limit $\epsilon \to 0$ are
\begin{equation}
\langle \phi_{T\pm}(x,t) \phi_{T\pm}(0,0)
\rangle=-i{\frac{\pi}{2}} {\rm sign}(xt) \label{eq:gffin}
\end{equation}
Using the above results we find that the propagator for an
operator $\Psi$ of the form of Eq.\ \ref{eq:opqp33}, in the limit
$x \to 0^+$, is
\begin{equation}
\langle \Psi^{\dagger}(0^+,t) \Psi(0,0)\rangle \propto
\;{\frac{1}{|t|^{g_t}}}\; e^{\displaystyle{-i{\frac{\theta}{2}}
{\rm sign}(t)}} \label{eq:qppropuu}
\end{equation}
where the exponent $g$ is
\begin{equation}
g_t=a_{C+}^2+a_{C-}^2 \label{eq:gqp}
\end{equation}
and $\theta$ is the statistical angle of Eq.\ \ref{eq:chst3}.

Below we will apply these results to the computation of tunneling
exponents to a number of FQH states of special interest. It is
worth to notice  that the exponent $g_t$ is determined by the
coefficients of the fields $\phi_{C\pm}$ alone, and that the
topological fields $\phi_{T\pm}$ only contribute to the
statistics. In addition, for $p=\pm 1$ the contribution of the
topological fields to the statistics of the excitations yields a
trivial multiple of $2\pi$. Hence, as anticipated above, for
$p=\pm 1$ the fields $\phi_{T\pm}$ drop out altogether. Finally,
by symmetry, the propagator for quasiparticles with spin down is
also given by Eq.\ \ref{eq:qppropuu}, with the same exponents.
Also by symmetry, the crossed propagator, which mixes up and down
quasiparticles, vanishes identically.

It is also an interesting question to ask which operator can
measure the {\sl relative statistics} of two quasiparticles. The
simplest way to do that is to add a boundary perturbation that
will allow for tunneling between different types of
quasiparticles. For example consider a perturbation well localized
at some point on the edge, with tunneling amplitude $\Gamma$. It
is easy to see that, to first order in $\Gamma$, a crossed
propagator is induced. This crossed propagator is just the square
of two propagators like Eq.\ \ref{eq:qppropuu} multiplied by a
phase factor of the relative statistics. In any event, just as in
the case of fractional statistics\cite{interferometer}, in
practice it will be necessary to consider an intereferometric
experiment (with two tunneling centers) in order to measure
relative statistics.

We will compute the tunneling exponents  in two cases of special
interest: the symmetric $SU(2)$ and $U(1) \times U(1)$ states. In
particular we will discuss different tunneling processes and give
the tunneling exponents. There are three situations of physical
interest: i) internal tunneling of quasiparticles, ii) tunneling
of electrons between {\sl identical} fluids, and iii) electron
tunneling between distinct fluids. Because these states are
symmetric, even though there are several tunneling channels
leading to a conductance matrix, it turns out that the {\sl
exponents} associated with different tunneling channels are equal
while the amplitudes in general are different. In what follows we
will only discuss the scaling exponents. These can be calculated
by a straightforward generalization of the arguments of ref.\
[\ref{ref:wen-tunnel}], based on the formalism of ref.\
[\ref{ref:ssw}], for the case of both internal tunneling and of
tunneling of electrons between identical fluids. The tunneling
current $I(V)$ at bias voltage $V$ for identical layers, has the
scaling form
\begin{equation}
I_{bc}(V) \propto M_{bc} \; V^\alpha \label{eq:tun-curr}
\end{equation}
 as discussed by Wen\cite{wen-tunnel} and by Kane and
Fisher\cite{kane-fisher} for Laughlin fluids. The  exponent
$\alpha$ is determined by the scaling dimension of the tunneling
operator. Here we have denoted the amplitudes for the different
channels by a $2 \times 2$  matrix $M_{bc}$ (which we will not
calculate).

For {\sl internal tunneling} of quasiparticles, the tunneling
exponent $\alpha_{\rm qp}$ is given in terms of the exponent of
the quasiparticle propagator $g_{\rm qp}$ by the formula
\begin{equation}
\alpha_{\rm qp}=2 g_{\rm qp}-1 \label{eq:alfaqp}
\end{equation}
Internal tunneling of quasiparticles is always a relevant
perturbation. Consequently, the system always flows at low
energies to a regime in which the fluid splits in
two\cite{kane-fisher}. Hence, in this regime quasiparticle
tunneling yields a reduction of the differential conductance from
its value at zero tunneling, the Hall conductance of the fluid. In
this regime Eq.\ \ref{eq:tun-curr} for the tunneling current holds
at high frequencies. In the case of {\sl electron tunneling}
between identical fluids Eq.\ \ref{eq:tun-curr} holds at low
frequencies, with an exponent $\alpha_e$  given by
\begin{equation}
\alpha_e=2g_e-1 \label{eq:alfae}
\end{equation}
where $g_e$ is the exponent of the electron propagator. Finally,
for tunneling of electrons between an external lead (a Fermi
liquid) and a bilayer FQH state, the scaling form Eq.\
\ref{eq:tun-curr}, also valid at low frequencies, has an exponent
$\alpha_t$ given by
\begin{equation}
\alpha_t=g_e \label{eq:alfa-diff}
\end{equation}
where $g_e$ is the exponent of the electron propagator of  the FQH
fluid\cite{chamon}. For the case of tunneling of electrons into a
single layer FQH state in the Jain sequence, this exponent is
equal to $1/\nu$ ( see ref.\ [ \ref{ref:ref1}]).

\subsection{ $SU(2)$ states.}

Here we give a description of the spectrum of  the edge states of
the $SU(2)$ FQH states. Recall that for these states $m=n+1$, and
the filling fraction is $\nu=2p/(2np+1)$. A special feature of the
$SU(2)$ states is that the chirality of the spin field is $s={\rm
sgn}(p)$. Hence, for the $SU(2)$ states only the sign of $p$
matters.

\begin{enumerate}
\item
{\sl Quasiholes}:\\ The operators that create quasiholes with spin
up or down are given in Eq.\ \ref{eq:opqp5}. For the $SU(2)$
states, the quantum numbers of these quasiholes are
\begin{eqnarray}
Q                   =&&{\frac{e}{2np+1}} \nonumber \\ S_z
=&&\pm {\frac{1}{2}} {\rm sign} (p) \nonumber \\
{\frac{\theta}{\pi}}=&& 1+{\frac{n}{2np+1}} \nonumber \\ &&
\label{eq:qn-qpsu2}
\end{eqnarray}
The exponents $g_{\rm qp}$ for quasiparticles of either spin are
\begin{equation}
g_{\rm qp}= \displaystyle{ \left\{
\begin{array}{cc}
\frac{n+\frac{1}{p}}{2np+1} & {\rm for} \;\; p>0 \\ {}
&  {}                \\ \frac{n}{2n|p|-1}           & {\rm for}
\;\; p<0
\end{array}
\right. } \label{eq:gqpsu2}
\end{equation}
Below we will use these results to compute the exponents for
internal tunneling.
\item
{\sl Bound States}:\\ Let us consider bound states of a
quasiparticle and a quasihole. Such states are electrically
neutral and have spin projection $\pm 1$.  For all symmetric FQH
states in bilayers, the operator that creates a bound state of two
quasiparticles, with total $S_z=\pm1$ and zero electric charge, up
to singular normalization factors (see Appendix \ref{app:AA}), are
\begin{eqnarray}
\lefteqn{S^\pm (0) \equiv \lim_{z \to 0} :\Psi_{{\rm
qp},\uparrow,\downarrow}^\dagger(z) \Psi_{{\rm
qp},\downarrow,\uparrow}(0): \propto} \nonumber \\ &&
e^{\displaystyle{ \pm \frac{i}{p} \sqrt{
\frac{2}{|m-1-n+\frac{1}{p}|} }\phi_{C-}(0) }} \;
e^{\displaystyle{\mp i \sqrt{\frac{2}{|p|}} \phi_{T-}(0) }}
+\ldots \nonumber \\ && \label{eq:S+}
\end{eqnarray}
For example, in the special case of the $SU(2)$ states we find
\begin{equation}
S^\pm(0) \propto e^{\displaystyle{\pm i \sqrt{\frac{2}{|p|}}
\phi_{T-}(0)}} e^{\displaystyle{\mp i \sqrt{\frac{2}{|p|}}
\phi_{C-}(0)}} \label{eq:Spmsu2}
\end{equation}
In addition to these operators, which have $S_z=\pm1$, we also
have the spin current operator $j_z$, given by Eq.\
\ref{eq:spin-current}, which has zero electric charge and $S_z=0$.
For $p=1$, the primary Halperin states, these three operators are
the local generators of a $su(2)_1$ Kac-Moody algebra. This
algebra, and its generators, can be used to construct the entire
Hilbert space of these spin edge states. However, for general $p
\neq 1$ this is not possible since the off-diagonal generators do
not have scaling dimension $1$. Thus, for general $p$, there is
only a global $SU(2)$ spin symmetry in the sense that the states
are arranged into $SU(2)$ multiplets. However, for $p=r^2$ (where
$r$ is an integer) there exist complexes of $r$ bound states, of
the type constructed above, which constitute the off-diagonal
generators of a Kac-Moody algebra $su(2)_1$.

Apart from these electrically neutral bound states, there are
charged bound states as well. For instance, let us consider bound
states of two quasiparticles. There are four operators that create
such charged bound states. In the Appendix \ref{app:AA} we show
that it is possible to arrange these four operators into linear
combinations that create  spin singlet pairs and three that create
spin triplet pairs. The spin singlet operator
\begin{eqnarray}
\Psi_{\rm singlet} \propto : &&e^{\displaystyle{-i
\frac{\sqrt{\nu}}{p} \phi_{C+}(z)}}: \; \; :\partial_z
\phi_{C-}(z):\nonumber \\ &&\times \; :
e^{\displaystyle{i2\sqrt{1+\frac{1}{2p}} \phi_{T+}(z)}} :
\nonumber \\ && \label{eq:singlet}
\end{eqnarray}
plays a central role in the construction of the electron operator.
Here too, the three spin triplet operators only have the same
scaling dimension for the primary Halperin FQH states, which have
$p=1$. Once again we see the same pattern: states with the correct
charge and spin quantum numbers fail to have the same scaling
dimension except for $p=1$.
 \item
{\sl Electrons}:\\ In order to construct an electron operator for
these FQH states we use the fact that they have filling fraction
$\nu={\frac {2p} {2np+1}}$ with $n$ even, and that both
quasiparticles have the same charge $Q=\frac {1}{2np+1}$.

Therefore we can obtain an object with charge $Q= -e$ if we
construct a composite object made of $2np+1$ quasiparticles. In
addition, in order to be an electron, this object must have total
spin $S=1/2$. Clearly, out of $2np+1$ objects we can construct a
number of states (operators) with different total spin. For
instance with two quasiparticles it is possible to construct spin
singlet states with total spin $S=0$, or triplet states with
$S=1$. Since, in order to make an electron we must construct an
object with total spin $S=1/2$, we take $np$ singlets constructed
as it is shown in Appendix \ref{app:AA}, and an extra
quasiparticle whose spin projection will determine the spin
projection of the electron operator. Following this prescription
the electron operator results
\begin{eqnarray}
\lefteqn{\Psi (x)_{{\rm e},\uparrow\downarrow}
 =e^{\displaystyle{- { \frac {i}{ \sqrt{\nu}}}\phi_{C+} \mp
{\frac{i}{\sqrt{2|p|}}} \phi_{C-} }}} \nonumber \\ &&\;\;\;\;\;\;
\times \; e^{\displaystyle{ i(2np+1) \; {\rm sign} (p)\; \sqrt{1+
\frac{1}{2p}} \phi_{T+} \pm { \frac {i}{\sqrt{2|p|}}} \phi_{T-} }}
\nonumber \\ && \label{eq:elprop}
\end{eqnarray}
up to irrelevant operators, whose form is discussed in Appendix
\ref{app:AA}. This operator creates states with charge $Q=-e$,
spin projection $S_z=\pm {\frac{1}{2}}$, and statistics $\theta=
\pi (2np+1)(2np+n+1)$. The exponent $g_e$ for the electron
operator is
\begin{equation}
g_e=
\left\{ 
\begin{array}{cc}
n+{\frac{1}{p}} & {\rm for} \; p>0\\
n             & {\rm for} \; p>0
\end{array}
\right.
\label{eq:ge}
\end{equation}
This result holds for both signs of $p$.

For the special case of the Halperin state $(3,3,2)$, with
$\nu=2/5$, the electron operator is
\begin{equation}
\Psi_{{\rm e},\uparrow,\downarrow}(x)= e^{\displaystyle{-i
{\sqrt{\frac{5}{2}}} \; \phi_{C+} \mp {\frac{i}{\sqrt{2}}} \;
\phi_{C-}}} \label{eq:el5-2}
\end{equation}
where we have dropped the contributions of the topological fields
since for the $p=1$ states they act like the identity operator. 
The exponent for the electron propagator results $g_e=3$. In
Eq.\ \ref{eq:el5-2} we have kept only the most relevant operators
which contribute to the electron operator, and neglected
subleading irrelevant operators involving  $\partial_x  \exp
\left(-i{ \frac {1}{ \sqrt{2}}} \phi_{C-}\right)$. These
irrelevant operators appear in the operator product expansion with
well defined coefficients which are calculated in  Appendix
\ref{app:AA}.

Similarly, for the singlet state at $\nu=2/3$, which has $m=n+1=3$
and $p=-1$ (\ie\ it belongs to the reversed sequence), the electron operator is
\begin{equation}
\Psi_{{\rm e},\uparrow,\downarrow}(x)= e^{\displaystyle{-i
{\frac{3}{\sqrt{2}}} \; \phi_{C+} \mp {\frac{i}{\sqrt{2}}} \;
\phi_{C-}}} \label{eq:el2-3}
\end{equation}
 Notice that in this case the electron can be viewed as a bound state of a  of a right 
moving electron and a right moving semion of the charge sector, and a left moving semion
of the spin sector. Also, although the
total exponent of the electron is in this case
$g_e=\frac{3}{2}+\frac{1}{2}=2$, since the velocities of the
charge and spin bosons are different, the electron propagator
formally still has a branch cut.
\end{enumerate}

We will now apply the results derived above to the computation of
the tunneling exponents for the $SU(2)$ states.

\begin{enumerate}
\item
{\sl Internal quasiparticle tunneling}:\\ For the $SU(2)$ states
the exponent of the quasiparticle propagator is  calculated in
Eq.\ \ref{eq:gqpsu2}. In particular, for the primary Halperin
state $(3,3,2)$ we find the exponent $g_{\rm qp}= {\frac {3}{5}}$.
In this regime, the two-terminal differential conductance is
reduced from the quantized bulk Hall conductance due to tunneling
of quasiparticles. This reduction  has the scaling form
$V^{\alpha_{\rm qp}-1}$, with $\alpha_{\rm qp}$ given by Eq.\
\ref{eq:alfaqp}. In particular, for the $(m,m,m-1)$ $SU(2)$
Halperin states we find the exponent $\alpha_{\rm qp}=1/(2m-1)$.
For the case of the reversed sequence states, with $p=-1$, the exponent is
obtained upon substituting $m \to m-2$. For the $2/5$ spin singlet
state the exponent is $g_{\rm qp}=3/5$, we get $\alpha_{\rm
qp}=1/5$, and  the two-terminal conductance follows the law
$G_{\rm qp} = {\frac 2 5}{\frac {e^2}{h}}- {\rm const.} \times
\left(T_K/V\right)^{4/5}$. Once again, this law holds only at
large voltages $V \gg T_{K}$, where $T_K$ is a non trivial
crossover energy scale qualitatively similar to the Kondo scale in
quantum impurity systems\cite{chamon}. By a similar calculation,
for the $\nu=2/3$ singlet state we find $g_{\rm qp}=2/3$ and an
exponent for the tunneling current of $\alpha_{\rm qp}=1/3$.
\item
{\sl Electron tunneling between identical }$SU(2)$ {\sl states}:\\
~From the exponent of the electron propagator of Eq.\ \ref{eq:ge},
we find that the exponent of the differential conductance for
tunneling of electrons between identical $SU(2)$ singlet FQH
states is
\begin{equation}
\alpha_e=
\left\{ 
\begin{array}{cc}
2n-1+{\frac{2}{p}}  & {\rm for} \; p>0\\
2n-1              & {\rm for} \; p>0
\end{array}
\right.
\label{eq:alfa-e-su2}
\end{equation}
In particular, for the $(m,m,m-1)$ primary states we find
$\alpha_e=2m-1$, for $p>0$, and $\alpha_e=2m-3$ for $p=-1$. For the
$2/5$ spin singlet state we find that the tunneling current of
electrons obeys the law $I \sim V^5$ and $G_e \sim V^4$. Instead,
for the spin singlet state at $\nu=2/3$ we get
$\alpha_e=3$, and the current tunneling obeys the 
law $I \sim V^3$.
 For the states in the second level of the hierarchy
which have already been observed experimentally \cite{cho47}, \ie,
for $4/7$ and $4/9$ the exponents are $\alpha_e =3$ and $4$
respectively.
\item
{\sl Electron tunneling into an} $SU(2)$ {\sl state from a Fermi
liquid}:\\ Finally, the exponent for the differential conductance
for tunneling into an $SU(2)$ singlet state from an external lead
is
\begin{equation}
\alpha_t=g_e=
\left\{ 
\begin{array}{cc}
m-1+{\frac{1}{p}} & {\rm for} \; p>0\\
m-1             & {\rm for} \; p>0
\end{array}
\right.
\label{eq:alpha-t-su2}
\end{equation}
Hence, for the $2/5$ state we get $I \sim V^3$ and $G_t \sim V^2$.
For $\nu=2/3$ we find that the tunneling current of electrons from
a Fermi liquid obeys the law $ I \sim V^2$ . For the
states $4/9$ and $4/7$ the current will obey $I \sim V^{5/2}$ and
$I \sim V^{2}$ respectively.

Quite generally, at the edge of an $SU(2)$ state, at filling
factor $\nu=2p/(2np+1)$, the tunneling density of states for
electrons obeys the law $|\omega|^{g_e- 1}$, where
$g_e$ is given in Eq.\ \ref{eq:alpha-t-su2}. In contrast, the corresponding spin polarized
state with the same filling factor has a tunneling density of
states for electrons with the law  $|\omega|^{{\frac {1}{\nu}}- 1}
$, see ref.\ [\ref{ref:ref1}].
\end{enumerate}

\subsection { $U(1) \times U(1)$ states.}

The $U(1) \times U(1)$ symmetric states have filling fraction
$\nu=2p/((m+n-1)p+1)$ with $n$ odd. Here we will consider the
states which are not in the $SU(2)$ subclass and are
incompressible. The special case of the $(m,m,m)$ states, \ie\
states with $m=n+2$ and $p=-1$, are actually the same as the
$(m,m,m)$ states and are not a separate case.

Here, unlike the special case of the $SU(2)$ states, there are two
general types of states: (a) states with $m>n+1$ , and (b) states
with $m<n$. For states with $m> n+1$, we have
$m-n-1+\frac{1}{p}>0$. Hence, in this case the charge field
$\phi_{C+}$ and the spin field $\phi_{C-}$ have the same
chirality, \ie\ $s=+1$. In the other case, $m<n$, the opposite
inequality holds, $m-n-1+\frac{1}{p}<0$. Hence $s=-1$, and the
chirality of the spin field is opposite to the chirality of the
charge field. The same is true for $m=n$ and $p \neq 1$.

The spectrum of the edge states of the $U(1) \times U(1)$ FQH
states consists of the following,
\begin{enumerate}
\item
{\sl Quasiholes}:\\ There are two quasiholes with the quantum
numbers
\begin{eqnarray}
Q=&&\frac{e}{(m+n-1)p+1}\nonumber \\ 
S_z=&&\pm {\frac{{\rm sgn}(p)}{2\left|\left(m-n-1\right)p+1\right|}} \nonumber \\
\frac{\theta}{\pi}=&&1+\frac{1}{p}-\frac{m-1+\frac{1}{p}}{\left((m-1)p+1\right)^2-n^2p^2}
\nonumber \\ && 
\label{eq:qnu1}
\end{eqnarray}
The general form of the operator that creates the quasiholes of  a
general $U(1) \times U(1)$ state was given in Eq.\ \ref{eq:opqp5}.
In particular, in the case of the $(3,3,1)$ state the operators
that creates quasiholes with both polarizations are much simpler,
\begin{equation}
\Psi_{\rm qp} \propto e^{\displaystyle{-\frac{i}{\sqrt{8}}
\phi_{C+}}} \; e^{\displaystyle{\pm \frac{i}{2} \phi_{C-}}}
\label{eq:qp-331}
\end{equation}
The quasiparticle propagators at the edge of a $U(1) \times U(1)$
state have exponents $g_{\rm qp}$ given by
\begin{equation}
g_{\rm qp}={\displaystyle{ \left\{
\begin{array}{cc}
{\displaystyle{\frac{m-1+\frac{1}{p}}{\left((m-1)p+1\right)^2-n^2p^2}}}
& {\rm for}\; s>0 \\ {} & {} \\
{\displaystyle{\frac{n}{n^2p^2-\left((m-1)p+1\right)^2}}} & {\rm
for}\; s<0
\end{array}
\right. }} \label{eq:gqpu1}
\end{equation}
\item
Neutral fermions:\\ There are both charged and neutral bound
states, and their statistics depends on which particular FQH state
is being discussed. Here we will consider only bound states for
the $(3,3,1)$ state which have a number of interesting features.
An operator quasiparticle-quasihole bound state with
$(k_1,k_2)=(1,-1)$. It has zero charge, spin $S_z=1/2$, and it is
a fermion. It is a neutral fermion operator created by
\begin{equation}
\Psi_{\rm NF, \uparrow, \downarrow}(0) \propto
e^{\displaystyle{\pm i \phi_{C-}}} \label{eq:NF331}
\end{equation}
This is the chiral Dirac fermion at the edge discussed by
Milovanovic and Read\cite{milovanovic}. The exponent for the
neutral fermion is $g_{\rm NF}=1$. The neutral fermion has
dimension $1/2$ and the corresponding tunneling operator has
dimension $1$. Hence, the operator that tunnels neutral fermions
at one point is {\sl marginal}.

The operators that create charge neutral quasiparticle bound
states with $S_z=\pm 1/2$  in a general symmetric $U(1) \times
U(1)$ FQH state with $m-n$ even and $p$ odd, have  the form
\begin{eqnarray}
\Psi_{{\rm neutral},\uparrow,\downarrow}=&& e^{\displaystyle{ \mp
i \sqrt{\frac{1}{2} |m-n-1+\frac{1}{p}|} \phi_{C-} }} \nonumber \\
&& \times \; e^{\displaystyle{\pm i \sqrt{\frac{|p|}{2}}
|m-n-1+\frac{1}{p}| \; \phi_{T-} }} \nonumber \\ && \label{eq:NO}
\end{eqnarray}
In particular, all of the primary $(m,m,m-2)$ states have a Dirac
fermion in their spectrum. The exponents for the neutral states
are
\begin{equation}
g_{\rm neutral}=\frac{1}{2} \left|m-n-1+\frac{1}{p}\right|
\label{eq:gN}
\end{equation}
Except for the case of the FQH states $(m,m,m-2)$, and their
descendants with $p$ odd, the processes of tunneling of neutral
$S_z=\pm 1/2$ excitations will turn out to be irrelevant. However,
in all cases where they are allowed, operators that represent the
leading processes with  spin flip {\sl without} tunneling of
charge will necessarily involve the neutral operators of Eq.\
\ref{eq:NO}.
\item
Spin singlet bound states:\\ Here we will discuss only the
$(3,3,1)$ primary state. Consider first a bound state with two
quasiholes. There are four such states: the singlet state, and the
three triplet states. These operators have
$(k_1,k_2)=(2,0),(0,2)$, and the symmetric and antisymmetric
combinations of the $(1,1)$ operator (see Appendix \ref{app:AA}):
\begin{eqnarray}
\Psi_{\pm 1/2}       \propto &&
e^{\displaystyle{-\frac{i}{\sqrt{2}}\phi_{C+}}}
                                e^{\displaystyle{\mp    i           \phi_{C-}}}
\nonumber \\ \Psi_0               \propto &&
e^{\displaystyle{-\frac{i}{\sqrt{2}}\phi_{C+}}} \nonumber \\
\Psi_{\rm singlet}   \propto &&
e^{\displaystyle{-\frac{i}{\sqrt{2}}\phi_{C+}}}
                      i \partial_z \phi_{C-}
\nonumber \\ && \label{eq:charged-bs}
\end{eqnarray}
These operators have dimensions $3/4$ (two states), $1/4$ (one
state) and $5/4$ (one state) respectively. The states they create
have charge $1/2$, $S_z=1/2$, and are semions. Their exponents are
$g=3/2,1/2,5/2$. Hence only the tunneling operator
$\Psi_{0,R}^\dagger \Psi_{0,L}$  is relevant (it has dimension
$1/2$). (Here $R,L$ here denote the edges of a FQH fluid.)
Finally, since the $(3,3,1)$ state is particle-hole symmetric (or
rather, it is compatible with it), the adjoint of the operators of
Eq.\ \ref{eq:charged-bs} create states with charge $-1/2$.

Next we consider bound states of four quasiparticles, \ie\ bound
states of the bound states. In particular the operators
\begin{eqnarray}
j_\pm \propto && \left(\Psi_0\right)^2 \propto
e^{\displaystyle{\pm i\sqrt{2}\phi_{C+}}} \nonumber \\ j_0
\propto && i \partial_z \phi_{C+} \nonumber \\ &&
\label{eq:su2-331}
\end{eqnarray}
create states with charge $1,0,-1$, $S_z=0$ and are bosons. These
operators span an $su(2)_1$ algebra of charge, a consequence of
particle-hole symmetry. All three states have dimension $1$ and
their exponents are $g=2$. Notice that in the $(3,3,2)$ we found
an $su(2)_1$ algebra for spin.
\item
{\sl Electrons}:\\ To construct a fermionic operator with charge
$Q= -1$ and the lowest spin projection in a general $U(1) \times
U(1)$ state, we need to take $(m+n-1)p+1$ quasiparticles arranged
in such a way that they have the required properties. It is simple
to check, following the prescriptions given in Appendix
\ref{app:AA}, that the electron operator that results is  given by
\begin{eqnarray}
\Psi_{e,\uparrow,\downarrow}\propto && e^{\displaystyle{-i
\sqrt{\frac{1}{2}\left(m+n-1+\frac{1}{p}\right)} \phi_{C+}}}
\nonumber \\ && \times e^{\displaystyle{\mp
i\sqrt{\frac{1}{2}\left|m-n-1+\frac{1}{p}\right|} \phi_{C-}}}
\nonumber \\ &&\times e^{\displaystyle{i {\rm sign} (p) \;
\sqrt{1+\frac{1}{2p}} \left(\left(m+n-1\right)p+1\right) \phi_{T+}
}} \nonumber \\ && \times e^{\displaystyle{i
\frac{\left(m-n-1\right)p+1}{\sqrt{2|p|}} \phi_{T-} }} \nonumber
\\ &&
\end{eqnarray}
where we have dropped an overall singular coefficient. The
exponent of the propagator for the electron operator is
\begin{equation}
g_e=m-1+\frac{1}{p} \label{eq:ge-electron}
\end{equation}
Notice that for the Halperin states, $p=1$ and $g_e=m$. For the
reversed sequence states with $p=-1$, we get $g_e=m-2$. In particular, in
the special case of the state $(3,3,1)$ the electron operators are
\begin{equation}
\Psi_{e,\uparrow,\downarrow}\propto e^{\displaystyle{-i\sqrt{2}
\phi_{C+} }} \; e^{\displaystyle{ \mp i \phi_{C-}}}
\label{eq:electron331}
\end{equation}
with an exponent $g_e=3$. Once again, there is no contribution
from the topological fields $\phi_{T \pm}$ to the electron
operator since here $p=1$.

Note that the electron operator of Eq.\ \ref{eq:electron331} can
be regarded as bound state of a neutral fermion $\exp(i\phi_{C-})$
and a charge $1$ boson $\exp(-i\sqrt{2} \phi_{C+})$.

\end{enumerate}

We will now apply the results derived above to the computation of
the tunneling exponents for the $U(1) \times U(1)$ states.

\begin{enumerate}
\item
{\sl Internal quasiparticle tunneling}:\\ For the $U(1) \times
U(1)$ states, the exponents for quasiparticle propagators were
calculated in Eq.\ \ref{eq:gqpu1}. In particular, for the primary
$(m,m,n)$ states, with $p=1$ and $m>n+1$, we get
\begin{equation}
g_{\rm qp}=\frac{m}{m^2-n^2}
\end{equation}
Thus, for the $(3,3,1)$ state the exponent is $g_{\rm qp}= {\frac
{3}{8}}$, the tunneling exponent is $\alpha_{\rm qp}=2g_{\rm
qp}-1=-1/4$, and the reduction of the conductance follows the law
$G_{\rm qp}  = {\frac 1 2}{\frac {e^2}{h}}- {\rm const.} \times
\left(T_K/V\right)^{1/4}$.

It is also possible to consider tunneling of composites of
quasiparticles. In any given state there are always several
tunneling processes that are relevant, although quasiparticle
tunneling is always the most relevant operator. In particular, in
the $(3,3,1)$ state, other relevant internal tunneling processes
involving the operator $\Psi_0$ (this process has dimension $1$),
as well as tunneling of neutral fermions, which is a marginal
operator. Thus the effect of a weak perturbation, such as a weakly
coupled gate, on an otherwise decoupled $(3,3,1)$ state is rather
complex. Nevertheless, one still expects that as the tunneling
matrix element grows bigger the system should flow to the weakly
coupled $(3,3,1)$ states, with a rather non-trivial crossover in
between.
\item
{\sl Electron tunneling between identical} $U(1)\times U(1)$ {\sl
states}:\\ The exponent for the electron propagator in all $U(1)
\times U(1)$ states is $m-1+\frac{1}{p}$. Thus, the exponent for
electron tunneling equals
\begin{equation}
\alpha_e=2\left({\displaystyle{m-1+\frac{1}{p}}}\right)-1
\end{equation}
In particular, for the $(3,3,1)$ state, the electron tunneling
exponent is $ g_{e}= 3 $, the tunneling current scales like $I_e
\propto V^5$, and the tunneling differential conductance follows
the law $G_e \propto V^4$. For all primary $(m,m,n)$ states,
$g_e=m$, the tunneling current behaves like $V^{2m-1}$ and the
differential conductance scales like $V^{2(m-1)}$.
\item
{\sl Electron tunneling into an} $U(1) \times U(1)$ {\sl state
from a Fermi liquid}:\\ Finally, we consider the case of tunneling
of electrons from a Fermi liquid into an edge state of an $U(1)
\times U(1)$ state. The tunneling current now scales with an
exponent $\alpha_t = g_e$. Thus, we find the general result
\begin{equation}
I_t \propto V^{\displaystyle{m-1+\frac{1}{p}}}
\end{equation}
In particular, the tunneling current into a $(m,m,n)$ state scales
like $V^m$, and the differential conductance scales like
$V^{m-1}$. Finally, we note that the tunneling density of states
for electrons into a general $U(1) \times U(1)$ edge as a function
of frequency $\omega$ scales like $|\omega |^{m-2+{\frac
{1}{p}}}$.
\end{enumerate}

\section{ Edge theory for $(m,m,m)$ states}
\label{sec:mmm-edge}

In section \ref{sec:mmm} we derived an effective theory for the
bulk $(m,m,m)$ states. The effective Lagrangian of this theory,
Eq.\ \ref{eq:Leff-mmm}, is a sum of two decoupled terms: (i) a
charge sector $b_\mu^+$ with an effective Lagrangian identical to
that of a single layer Laughlin state at filling factor $\nu=1/m$,
and (ii) a $2+1$-dimensional Maxwell-like Lagrangian for the
neutral sector, which is dual to the Lagrangian of a (massless)
phase field $\theta$ Eq.\ \ref{eq:Ltheta}. In that section we
showed that the $(m,m,m)$ states have an $m$-fold topological
degeneracy  on the torus and constructed its excitation spectrum.
In this section we will construct the  theory of the edge states
for the $(m,m,m)$ states by means of  a line of reasoning
analogous to what we did for the other FQH states in bilayers.

The decoupling of the effective low energy theory means that the
correlation functions in the bulk are products of a factor for the
charge sector and a factor for the neutral sector. In principle we
expect the same factorization (``separation") to take place on the
edges as well. The only caveat here is that, while the charge
sector has a simple edge structure, identical to that of the
Laughlin single layer states constructed by Wen\cite{wen-edge},
the neutral sector has gapless excitations in the bulk and as such
its Hilbert space does not project to the edge. In other words,
the effective theory of the edge, if it exists at all, is not a
local chiral conformal field theory. That is not chiral is obvious
since the Maxwell theory is not chiral. In the case of the
Laughlin states, the conformal invariance of the edges is a
consequence of the bulk being an incompressible chiral topological
fluid. While the charge sector does satisfy these requirements,
the neutral sector does not. We will see that nevertheless a
theory of the edge can be constructed, but it is neither chiral
nor local.

\subsection{The charge sector of the edge $(m,m,m)$ states.}
\label{sec:charge-mmm}

Since the charge sector is identical to that of a Laughlin state
for a single layer system, the effective Lagrangian for the edge
states for the charge sector of the $(m,m,m)$ states is just a
theory of a chiral bose field $\phi_c$, \ie
\begin{equation}
{\cal L}_{\rm charge}[\phi_c]={\frac{m}{4\pi}} \left[\partial_x
\phi_c \partial_t \phi_c-v_c \left(\partial_x \phi_c\right)^2
\right] \label{eq:Lcharge-mmm}
\end{equation}
where we have taken the edge to be a straight line along the $x$
axis, and $v_c$ is the velocity of the edge of the charged
sector. As usual $v_c$ depends on the confining electric field and
on the interactions.

In a single layer FQH Laughlin state the chiral edge boson
$\phi_c$ would obey the periodicity condition
\begin{equation}
\phi_c(x+2\pi R,t)=\phi_c(x,t)+2 \pi R n
\label{eq:laughlin-compactification}
\end{equation}
which follows from the single-valuedness of the electron state.
Here $R=\sqrt{\nu}=1/\sqrt{m}$ is the so-called compactification
radius. It follows\cite{wen-edge} that the only allowed states are
the Laughlin quasiparticle $V_{qp}=\exp((i/\sqrt{m}) \phi_c)$ and
the electron operator $V_e=\exp(i{\sqrt{m}}\phi_c)$. However, in
the case of the $(m,m,m)$ states the Hilbert space is larger.
Indeed, in addition to the analogs of the Laughlin quasiparticle
and electrons, it is possible to construct more states by gluing
{\sl multivalued} (twisted) operators of the charge sector with
twisted states from the neutral sector. These states will be the
projection of the bulk vortices discussed in section \ref{sec:mmm}
to the edge. The simplest of these states is created by the
operator $V_{1/2} \sim \exp((i/(2 \sqrt{m})) \phi_c)$ which has
charge $e/(2m)$ and statistics $\pi/(4m)$. This state is made
consistent (\ie single valued) by a contribution from the neutral
sector that we will discuss below.

\subsection{The neutral sector of the edge $(m,m,m)$ states.}
\label{sec:neutral-mmm}

The effective edge theory of the neutral sector $\theta$ on a
region $\Omega$ is constructed as follows. We begin by demanding
that the total neutral current should vanish at the boundary
$\partial \Omega$. This condition implies that the field $\theta$
must obey {\sl von Neumann} boundary conditions at $\partial
\Omega$. Let us denote by $\phi_n$ the restriction of the bulk
neutral field $\theta$ to the boundary $\partial \Omega$, \ie ~
${\displaystyle{\theta|_{\partial \Omega}=\phi_n}}$, and enforce
this condition in the path-integral of the neutral sector by
writing
\begin{equation}
1=\int {\cal D} \phi_n \; \delta\left(\theta|_{\partial
\Omega}-\phi_n\right)= \int {\cal D} \phi_n {\cal D} \omega \;
e^{\displaystyle{i \int_{\partial \Omega}\omega
\left(\theta-\phi_n\right)}} \label{eq:one}
\end{equation}
The partition function (in imaginary time) $Z_{\rm neutral}$ of
the neutral sector on $\Omega$ with Neumann  boundary conditions
on $\partial \Omega$ takes the form
\begin{eqnarray}
\lefteqn{Z_{\rm neutral}=\int {\cal D} \phi_n {\cal D} \omega
{\cal D} \theta\; } \nonumber \\ &&
e^{\displaystyle{-{\frac{g}{8\pi^2}}\int_\Omega d^3x \;
\left(\partial_\mu \theta \right)^2+i \int_{\partial \Omega} d^2x
\omega \; \left( \theta - \phi_n \right) }} \nonumber \\ &&
\label{eq:Zn1}
\end{eqnarray}
After some elementary algebra and by making use of the Neumann
boundary condition the bulk neutral field can be integrated out to
give  an expression of the partition function in terms of the
boundary fields $\phi_n$ and $\omega$,
\begin{eqnarray}
\lefteqn{Z_{\rm neutral}=\int {\cal D} \phi_n {\cal D} \omega \;
e^{\displaystyle{-i \int_{\partial \Omega} d^2x \omega(x)
\phi_n(x)}} } \nonumber \\ \times \; &&
e^{\displaystyle{-{\frac{2\pi^2}{g}} \int_{\partial \Omega} d^2x
\int_{\partial \Omega} d^2x' \; \omega(x) \; G(x-x')|_{\partial
\Omega}\; \omega(x')}} \nonumber \\ && \label{eq:Zn2}
\end{eqnarray}
where $G(x-x')$ is the Green function on $\Omega$ satisfying
Neumann boundary conditions on $\partial \Omega$, \ie
\begin{equation}
-\left[{\frac{1}{v}} \partial_0^2 + v {\vec \nabla}^2 \right]\;
G(x-x')= \delta^3(x-x') \label{eq:GF}
\end{equation}
with the Neumann boundary condition,
\begin{equation}
\left. \partial_n G \right|_{\partial \Omega}=0 \label{eq:neumann}
\end{equation}
where $n$ is the direction normal to the boundary $\partial
\Omega$. It is straightforward to compute this Green function for
a straight edge.

We can now integrate out the field $\omega$ to find
\begin{equation} Z_{n}=
\int {\cal D} \phi_n \; e^{\displaystyle{-{\frac{g}{2 \pi^2}}
\int_{\partial \Omega}  \int_{\partial \Omega} \;
{\frac{\partial_\mu \phi_n(x) \partial_\mu \phi_n(x')}{4\pi
|x-x'|}}}} \label{eq:Zn4}
\end{equation}
where $\mu=0,1$. To simplify the notation we have dropped the
explicit dependence on the velocity $v$. In deriving Eq.\
\ref{eq:Zn4} we have also made use of the fact that, for a
straight edge, the Neumann Green function restricted to the edge
is twice the Green function in free space.

The effective action for the edge neutral field $\phi_n$ is
non-local and not chiral. This expression is a generalization of
the familiar Caldeira-Leggett effective action or a point-like
degree of quantum mechanical freedom coupled to an extended system
\cite{leggett}. More recently, Castro Neto, Chamon and Nayak
considered a generalization of the Caldeira-Leggett problem to the
case of an open Luttinger liquid, a Luttinger liquid coupled to a
higher dimensional massless field\cite{open}. Although similar in
spirit the details and assumptions of ref.\ [\ref{ref:open}] lead
to a form of the effective action different from Eq.\
\ref{eq:Zn4}. From a physical point of view, the non-locality of
the effective action of Eq.\ \ref{eq:Zn4} simply means that in a
system with massless bulk excitations there is no separation
between edge and bulk, and that edge excitations leak into the
bulk.

The form of the effective action for the neutral edge field
$\phi_n$, Eq.\ \ref{eq:Zn4}, has a number of important
consequences. It is easy to show that the vacuum expectation value
of Wilson loop operators infinitesimally close to the boundary (by
a small distance $a$) decay exponentially fast (in imaginary
time). This results follows as a consequence of the non-locality
of the action or, equivalently, from the existence of the massless
bulk mode. Thus, any excitation that carries a non vanishing
polarization (neutral charge) acquires an exponentially decaying
factor from its neutral sector, even if the charge sector alone
yields the familiar power law behavior of conformal field theory.
The decay length depends on both the coupling constant $g \propto
1/\ell_0$, and the small distance $a$. In any case $a$, which is
determined by the spatial extent of the bound state, for typical
pair interactions also scales with the magnetic length $\ell_0$.
Therefore, the characteristic decay length is of the order of the
magnetic length itself (up to a non-universal numerical constant).
This exponential decay takes place in both periodic and twisted
sectors of the field $\phi_n$. Thus, any bulk state with non-zero
polarization becomes a massive excitation at the edge.

On the other hand, states with zero polarization in the bulk
remain massless at the edge. This happens because the operator
$\exp(i \phi_n)$, which creates a unit of flux quanta at $x$ at
the boundary, has the correlation function
\begin{equation}
\langle e^{i\phi_n(x)} e^{-i\phi_n(x')}\rangle \propto
e^{\displaystyle{{\frac {\ell_0}{{\sqrt{2\nu} |x-x'|}}}}}
\label{eq:bcf}
\end{equation}
where we have set $|x-x'|^2\equiv v^2 |x_0-x_0'|^2+|x_1-x_1'|^2$.
This correlation function has no effect at long distances and/or
long times (up to small corrections). Thus the states with zero
polarization are the low energy Hilbert space.

We can now summarize the spectrum of edge states for the $(m,m,m)$
states.
It contains the Hilbert space of edge states of the single layer
Laughlin states at filling factor $\nu=1/m$. The only change is
that the operator that creates and electron in a state with zero
polarization now requires a factor that creates a flux quantum of
the neutral gauge field $b_\mu^-$. Thus there are two types of
electron operators
\begin{equation}
\psi_{e,\pm}^\dagger(x) \propto e^{\displaystyle{i \sqrt{m}
\phi_c(x)}} \; e^{\pm i\phi_n(x)} \label{eq:electron-neutral}
\end{equation}
where the sign $\pm$ indicates the sign of the charge that couples
to the gauge field $b_{\mu}^-$. Their correlation
functions are
\begin{equation}
\langle \psi_{e,+}^\dagger(x_0,x_1) \psi_{e,+}(x_0',x_1')\rangle
\propto {\frac{1}{(z-z')^m}}  \times \; e^{\displaystyle{{\frac{
\ell_0}{\sqrt{2\nu} |x-x'|}}}} \label{eq:neutral-electron-gf}
\end{equation}
where $z=x_1+i v_c x_0$. This result implies that the tunneling
density of states to the edge of an $(m,m,m)$ state is the same as
in the Laughlin state with the same filling factor, up to
corrections which are analytic in the frequency, \ie
$|\omega|^{m-1}+ {\cal O}(|\omega|^m)$. For all practical purposes
this contribution is a negligibly small effect.

In contrast, the propagators for the naively defined single-layer electron operators
decay exponentially fast with distance (and imaginary time). Consequently, the tunneling density of
states into an edge electron of a given layer
will show an energy gap for frequencies low compared with a
crossover energy scale, of the order of the cyclotron frequency,
and the usual power law at higher
frequencies, assuming that at these frequencies these states are still well defined
(which is unlikely). Bulk vortex states exhibit similar exponential decays at the
boundary.

It is interesting to use this edge theory to reconstruct the wave function of the $(m,m,m)$
states. Following Read and Moore \cite{moore-read} we can compute the
ground state wave function of the $(m,m,m)$ states. One has to
compute the correlation functions of $N/2$ electron operators of
the form $\psi_{e,+} \propto e^{\displaystyle{i \sqrt{m}
\phi_c(z)}} \; e^{ i\phi_n(z)}$ for $z=z_1,...,z_{N/2}$, and $N/2$
electron operators of the form $\psi_{e,-} \propto
e^{\displaystyle{i \sqrt{m} \phi_c(z)}} \; e^{ -i\phi_n(z)}$ for
$z=w_1,...,w_{N/2}$, times a neutralizing background for the
charged sector. Here the $z's$ and $w's$ are the coordinates
of the electrons on each layer in complex
notation.
 It is simple to check that this procedure gives the same
expression for the wave function as the one we derived in reference \cite{paper2}
\breakon
\begin{eqnarray}
&&
\Psi (z_1,...,z_{N/2},w_1,...,w_{N/2})
=
{\prod_{i<j=1}^{N/2}}
\left(z_i -z_j\right)^m
 {\prod_{i<j=1}^{N/2}}
 \left(w_i -w_j\right)^m \;
     \prod_{i=1}^{N/2}\prod_{j=1}^{N/2} \left(z_i -w_j\right)^{m}
   \nonumber \\
&&   e^{\displaystyle{-  {B\over 4} ({\sum_{i=1}^{N/2}}|z_i|^2
            + {\sum_{i=1}^{N/2}}|w_i|^2 )}}
         \;\times \;
 e^{\displaystyle{  -{\ell_0  \over { {\sqrt{ 2\nu}}}} \left(
{\sum_{i < j=1}^{N/2}} {1\over |z_i -z_j|}+
{\sum_{i < j=1}^{N/2}} {1\over |w_i -w_j|}-
{\sum_{i , j=1}^{N/2}} {1\over |z_i -w_j|} \right) }}
\nonumber \\
&&
\label{eq:wf8}
\end{eqnarray}
\breakoff
 This wave function differs from the conventional wave function for the $(m,m,m)$ state
 by the last
 factor. This extra factor has the same form as the Jastrow
 factor of the wave function for a superfluid. Here , as in the case of a suprefluid,
 this factor is
 due to the contribution of the ``phonons", the linearly dispersing Goldstone
 boson\cite{feynman}.

These results imply that tunneling experiments into the
(unreconstructed) edge states of an $(m,m,m)$ bilayer state are
likely to yield results analogous to the tunneling experiments
into a single layer Laughlin state\cite{chang1} provided that
there is a significant tunneling matrix element with the zero
polarization electron state. Such tunneling experiments have not
yet been done. However, recently I.\ B.\ Spielman and
coworkers\cite{spielman} have reported experiments of uniform tunneling into the bulk of
a $(1,1,1)$ state which show a
strongly  resonant tunneling conductance. Early on Wen and Zee predicted that the
gapless neutral mode would make resonant tunneling into the bulk
possible\cite{wenzee-gapless}.

\section{Conclusions}
\label{sec:conclusions}

In this paper we have presented a theory for the edge states for
spin polarized  bilayer and spin-$1/2$ single layer FQHE systems. We
assumed that the edge is sharp, unreconstructed and clean.

We began by constructing the simplest possible theory for bulk
states, compatible with the requirement of global gauge
invariance, with the correct topological degeneracy on a torus,
and with the smallest number of fundamental quasiparticles. Later
on, we used this bulk theory to find the physics of its edge
states. We found that the minimal edge theory thus derived has two
propagating fields that couple to the external sources and
represent the charge mode in each layer, and two topological
non-propagating fields. These latter fields play the role of Klein
factors, providing the right statistics for all the physical
operators.

We studied in detail all the Jain-like states for these systems,
whose primary states are the Halperin $(m_1,m_2,n)$ ones. In
particular, we described the spectrum of operators for the
symmetric states, \ie\ the $SU(2)$ states (the descendants of the
$(m,m,m-1)$ states), where the layer index is regarded as the spin
index, and the general $U(1) \times U(1)$ states. In all these
cases we explicitly constructed the operators that create the
quasiparticles (and quasiholes), charged and neutral bound states
(including neutral fermionic states) and the electron operators.
For the case of the $SU(2)$ states we showed how the symmetry is
realized in the spectrum and how it is promoted to a local
$su(2)_1$ spin current algebra.

We also calculated the propagators for the physical excitations.
We showed that the charge operators determine the exponent of the
time dependence of the propagators, and that the topological
operators only contribute to the statitics. For the primary
Halperin states, which have $p= \pm 1$, the contribution of the
topological operators to the statistics is a multiple of $2\pi$,
and thus can be ignored.

Afterwards we applied these results to compute the tunneling
exponents for all cases in three different situations: internal
tunneling of quasiparticles, tunneling of electrons between
identical liquids and tunneling of electrons into a FQH fluid from
an external Fermi liquid lead. As a general rule, we found  that
although the tunneling exponents are universal, in general they
are not equal to the inverse of the  filling factor.

In particular  we computed the tunneling exponents for the spin
singlet states that have been  observed experimentally
\cite{singlet-23}, whose filling fractions are $\nu =
2/3,2/5,4/7,4/9$. For instance , for electron tunneling between
two $2/3$ states we found that the tunneling current  obeys Ohm´s
law. The same result is obtained  for tunneling of electrons into
the $2/3$ state from a Fermi liquid. In general, at the edge of an
$SU(2)$ state, at filling factor $\nu=2p/(2np+1)$, the tunneling
density of states for electrons obeys the law $|\omega|^{g_e- 1}$,
where $g_e=n+\frac{1}{p}$. In contrast, the corresponding spin
polarized state with the same filling factor has a tunneling
density of states for electrons with the law  $|\omega|^{{\frac
{1}{\nu}}- 1} $ (see ref.\ [\ref{ref:ref1}]). Experiments of
tunneling of electrons into the edge of spin singlet systems have
not been done yet although they should be possible in higher
density samples. Experiments of this type would be very useful to
sort out the subtle correlations that give rise these FQH states.

Finally, we presented a theory of the edges for the $(m,m,m)$
states. In this case there is a bulk Goldstone boson which affects
the edge physics. The effective theory of the edges of the
$(m,m,m)$ states is a chiral boson for the charge mode (with the
same compactification radius as the Laughlin states), and a non
local (both in space and time) non-chiral theory for the neutral
mode. The non-locality of the neutral sector is due to the
existence of a massless Goldstone mode in the bulk. Physically, it
means that in a system with massless bulk excitations there is no
separation between edge and bulk, and that the edge excitations
leak into the bulk. For the $(m,m,m)$ states the operator that
creates an electron in a state with zero polarization  is the same
as the one for the Laughlin states at filling factor $\nu=1/m$,
times a factor that creates a flux quantum of the neutral gauge
field. This last factor will give a negligible contribution to the
tunneling density of states. Therefore, tunneling experiments into
the edges of an $(m,m,m)$ state should give  the same results as
for the corresponding single layer Laughlin state, provided there
is a sufficiently large overlap with the unpolarized electron
state.

\section{Acknowledgements}

This work was supported in part by the NSF grant NSF DMR98-17941
at the University of Illinois at Urbana-Champaign (EF), and by
CONICET,  Fundaci\'on Antorchas, and ANPCyT (pict 98/03-03924)
(Argentina) (AL).

\appendix
\section{Operator Product expansions }
\label{app:AA}

In this Appendix we calculate the Operator Product Expansions
(OPE's) for the quasiparticle operators of the symmetric bilayer
states. These OPE's apply for the edge states of both the $SU(2)$
and the $U(1) \times U(1)$ FQH states.

In order  to compute products of quasiparticle operators Eq.\
\ref{eq:opqp5} of the form $\Psi (x)_{{\rm qp},\uparrow} \Psi
(y)_{{\rm qp},\downarrow}$ or $\Psi (x)_{{\rm qp},\uparrow} \Psi
(y)_{{\rm qp},\uparrow}$, we use the following result for the
OPE's of two vertex operators\cite{cft} \breakon
\begin{equation}
: e^{i\beta \phi(x)}: \; : e^{i\beta' \phi(y)}: = :e^{i[\beta
\phi(x) +\beta' \phi(y)]}: e^{- \beta \beta' \langle  \phi(x)
\phi(y) \rangle}
\end{equation}
Hence, the product of two different quasiparticle operators is
\begin{eqnarray}
\Psi _{{\rm qp},\uparrow}(x)\Psi _{{\rm qp},\downarrow}(y)=&&
\left( {\frac {|x-y|}{a_0}}\right)^{\displaystyle{
{\frac{-n}{p^2\left(\left(m-1+\frac{1}{p}\right)^2-n^2\right)}}}}
\; e^{\displaystyle{\langle \phi_{T}(x) \phi_{T}(y)\rangle}}
\nonumber \\ && \times \;
:e^{\displaystyle{-i{\frac{\sqrt\nu}{2p}}(\phi_{C+}(x)+\phi_{C+}(y)
)}}: \;
:e^{\displaystyle{i{\frac{1}{p\sqrt{2(m-n-1+\frac{1}{p})}}}(\phi_{C-}(x)-\phi_{C-}(y))}}:
\nonumber \\ && \times \;
:e^{\displaystyle{i\sqrt{1+\frac{1}{2p}}(\phi_{T+}(x)+\phi_{T+}(y))}}:
\;
:e^{\displaystyle{i{\frac{1}{\sqrt{2p}}}(\phi_{T-}(x)-\phi_{T-}(y))}}:
\nonumber \\ &&
 \label{eq:qp34}
\end{eqnarray}
and the product of two identical quasiparticle operators results
in the following expansion
\begin{eqnarray}
\Psi _{{\rm qp},\uparrow}(x)\Psi _{{\rm qp},\uparrow}(y)= &&
\left({\frac{|x-y|}{a_0}}\right)^{\displaystyle{ {\frac
{m-1+\frac{1}{p}}{p^2((m-1+\frac{1}{p})^2-n^2)}}}} \;
e^{\displaystyle{(1+\frac{1}{p})\langle \phi_{T+}(x)
\phi_{T+}(y)\rangle}} \nonumber \\ && \times \;
:e^{\displaystyle{-i{\frac{\sqrt\nu}{2p}}(\phi_{C+}(x)+\phi_{C+}(y)
)}}:
:e^{\displaystyle{i{\frac{1}{p\sqrt{2(m-n-1+\frac{1}{p})}}}(\phi_{C-}(x)+\phi_{C-}(y))}}:
\nonumber \\ && \times \;
:e^{\displaystyle{i\sqrt{1+\frac{1}{2p}}(\phi_{T+}(x)+\phi_{T+}(y)
)}}:
:e^{\displaystyle{-i{\frac{1}{\sqrt{2p}}}(\phi_{T-}(x)+\phi_{T-}(y))}}:
\nonumber \\ && \label{eq:qp35}
\end{eqnarray}
To construct the singlet and triplet states we need to compute the
following products of operators, in the limit $y \rightarrow
(x-a_0)$,
\begin{eqnarray}
\Psi _{{\rm qp},\uparrow}(x)\Psi _{{\rm qp},\downarrow}(y) \pm
\Psi _{{\rm qp},\uparrow}(y)\Psi _{{\rm qp},\downarrow}(x)\;
\approx && :e^{\displaystyle{-i{\frac{\sqrt\nu}{p}}\phi_{C+}(x)}}:
:e^{\displaystyle{i2\sqrt{1+\frac{1}{2p}}\phi_{T+}(x)}}: \nonumber
\\ && \times \;
\left[:e^{\displaystyle{i{\frac{1}{p\sqrt{2(m-n-1+\frac{1}{p})}}}
(x-y) \partial \phi_{C-}(x)}}:
:e^{\displaystyle{i{\frac{1}{\sqrt{2p}}} (x-y) \partial
\phi_{T-}(x)}}: \right. \nonumber \\ && \left. \pm
:e^{\displaystyle{i{\frac{1}{p\sqrt{2(m-n-1+\frac{1}{p})}}} (y-x)
\partial \phi_{C-}(x)}}: :e^{\displaystyle{i{\frac{1}{\sqrt{2p}}}
(y-x) \partial \phi_{T-}(x)}}: \right] \nonumber \\ &&
\label{eq:qp346}
\end{eqnarray}
where we have ignored the amplitude factors. Since the topological
operators do not propagate $:\partial \phi_{T-}
e^{\displaystyle{i\alpha \phi_{T-}}}: =0$, therefore the operator
$:e^{\displaystyle{i{\frac{1}{\sqrt{2p}}} (x-y) \partial
\phi_{T-}(x)}}:$ acts like the identity operator. Hence, upon
expanding the exponentials in powers of $(x-y)$,  Eq.\
\ref{eq:qp346} becomes
\begin{eqnarray}
\lefteqn{\Psi _{{\rm qp},\uparrow}(x)\Psi _{{\rm
qp},\downarrow}(y) \pm \Psi _{{\rm qp},\uparrow}(y)\Psi _{{\rm
qp},\downarrow}(x) \approx
:e^{\displaystyle{-i{\frac{\sqrt\nu}{p}}\phi_{C+}(x)}}:
:e^{\displaystyle{i2\sqrt{1+\frac{1}{2p}}\phi_{T+}(x)}}: }
\nonumber \\ &&  \times \; \left[ 1+ i
{\frac{1}{p\sqrt{2\left(m-n-1+\frac{1}{p}\right)}}} (x-y) \partial
\phi_{C-}(x) \pm \left( 1-i
{\frac{1}{p\sqrt{2(m-n-1+\frac{1}{p})}}} (x-y) \partial
\phi_{C-}(x) \right) \right] \nonumber \\ && \label{eq:qp3466}
\end{eqnarray}
Analogously, one can calculate
\begin{equation}
\Psi _{{\rm qp},\uparrow}(x)\Psi _{{\rm qp},\uparrow}(y) \approx
:e^{\displaystyle{-i{\frac{\sqrt\nu}{p}}\phi_{C+}(x)}}:
:e^{\displaystyle{i{\frac{2}{p\sqrt{2(m-n-1+\frac{1}{p})}}}\phi_{C-}(x)}}:
:e^{\displaystyle{i2\sqrt{1+\frac{1}{2p}}\phi_{T+}(x)}}:
:e^{\displaystyle{-i{\frac{1}{\sqrt{2p}}} \phi_{T-}(x)}}:
\label{eq:qp349}
\end{equation}
in the limit $x \rightarrow y$.

We now apply the above results for the $SU(2)$ states. In this
case the singlet and triplet (with $S_z=0$) operators, $\Psi
_{singlet}$ and $\Psi _{0}$ respectively, can be written  as
follows:
\begin{eqnarray}
&&\Psi _{singlet}(x) \approx
:e^{\displaystyle{-i{\frac{\sqrt\nu}{p}}\phi_{C+}(x)}}:
:e^{\displaystyle{i2\sqrt{1+\frac{1}{2p}}\phi_{T+}(x)}}: i
\partial \phi_{C-}(x) \nonumber \\ && \Psi _{0}(x) \approx
:e^{\displaystyle{-i{\frac{\sqrt\nu}{p}}\phi_{C+}(x)}}:
:e^{\displaystyle{i2\sqrt{1+\frac{1}{2p}}\phi_{T+}(x)}}: \nonumber
\\ && \label{eq:qp3467}
\end{eqnarray}
respectively. The triplet operator, with $S_z=1$, is given by Eq.\
\ref{eq:qp349}.

Notice that the operator $\Psi _{0}$ and the one in Eq.\
\ref{eq:qp349} have the same dimension only if
\begin{equation}
\Delta (:e^{\displaystyle{-i\sqrt{\frac{2}{p}}\phi_{C-}(x)}}:) = 1
\label{eq:qp348}
\end{equation}
Since $\langle :e^{\displaystyle{-i\beta \phi(x)}}:
:e^{\displaystyle{-i\beta \phi(y)}}: \rangle = {\frac
{1}{|x-y|^{\beta ^2}}}$ and $ \Delta (:e^{\displaystyle{-i\beta
\phi(x)}}:)=\beta ^2 /2$, therefore we obtain
\begin{equation}
\Delta ( :e^{\displaystyle{-i\sqrt{\frac{2}{p}}\phi_{C-}(x)}}:)=
{\frac{1}{p}} \label{eq:qp3410}
\end{equation}
Therefore both operators have the same dimension only if $p=1$.
This is the only case where these three operators form a triplet.
In particular, the operators
\begin{equation}
E^\pm\equiv :e^{\displaystyle{\pm
i\sqrt{\frac{2}{p}}\phi_{C-}(x)}}: \;\;, \qquad H\equiv i\partial
\phi_{C-}(x) \label{eq:su2-1}
\end{equation}
generate the $su_1(2)$ algebra only if $p=1$, \ie\ only for the
state $(m,m,m-1)$ with $m$ odd, of which the state $(3,3,2)$ is a
particular case.

We discuss now  the leading irrelevant operators that contribute
to the electron operators. For instance for the $(3,3,2)$ state,
we have to compute the product of two singlets $\Psi _{singlet}(x)
$ and one quasiparticle operator $\Psi _{{\rm qp},\uparrow}(x)$.
The part corresponding to $\phi_{C-}(x)$ involves the OPE
\begin{equation}
 \left(\partial \phi_{C-} (z)\right)^2
:e^{\displaystyle{{\frac {i}{\sqrt{2p}}}\phi_{C-}(w)}}: \approx
{\frac{\Delta }{(z-w)^2}}
:e^{\displaystyle{{\frac{i}{\sqrt{2p}}}}\phi_{C-}(w)}:
+{\frac{\partial_w e^{\displaystyle{{\frac
{i}{\sqrt{2p}}}}\phi_{C-}(w)}}{z-w}} \label{opop1}
\end{equation}
where
\begin{equation}
\Delta = {\frac {1}{4p}} \label{eq:Delta}
\end{equation}

Finally we compute the spin flip operator,
\begin{eqnarray}
\Psi^{\dag} _{{\rm qp},\uparrow}(x)\Psi _{{\rm qp},\downarrow}(y)
=&& \left( {\frac{|x-y|}{a_0}}\right)^{\displaystyle{ {\frac
{m-1}{ p^2(2m-1)}}}} :e^{\displaystyle{\langle \phi_{T+}(x)
\phi_{T+}(y)\rangle}}:
:e^{\displaystyle{i{\frac{\sqrt\nu}{2p}}\left(\phi_{C+}(x)-\phi_{C+}(y)\right)-
{\frac{i}{p\sqrt{2}}}\left(\phi_{C-}(x)+\phi_{C-}(y) \right)}}:
\nonumber \\ && \times \;
:e^{\displaystyle{-i\sqrt{1+\frac{1}{2p}}\left(\phi_{T+}(x)-\phi_{T+}(y)
\right)}}:
:e^{\displaystyle{{\frac{i}{\sqrt{2p}}}\left(\phi_{T-}(x)+\phi_{T-}(y)
\right)}}: \nonumber \\ &&
 \label{eq:qp40}
\end{eqnarray}
In the limit  $y \rightarrow (x-a_0)$, the spin flip operator
becomes
\begin{equation}
\Psi^{\dag} _{{\rm qp},\uparrow}\Psi _{{\rm qp},\downarrow}\approx
:e^{\displaystyle{-i{\frac{\sqrt{2}}{p}}\phi_{C-}(x) }}:
:e^{\displaystyle{i\sqrt{\frac{2}{p}}\phi_{T-}(x)}}:
 \label{eq:qp41}
\end{equation}
\breakoff

\end{multicols}

\end{document}